\documentclass[aps,prfluids,preprint,groupedaddress]{revtex4-2}

\usepackage{graphicx}
\usepackage{newtxtext}
\usepackage{newtxmath}
\usepackage{natbib}
\usepackage{hyperref}
\usepackage{upgreek}
\usepackage{amsmath}
\usepackage{lineno}
\usepackage{bigdelim}
\hypersetup{
    colorlinks = true,
    urlcolor   = blue,
    citecolor  = black,
}

\newcommand{\RtoSLow}{R-to-S1}

\newcommand{\RtoSHigh}{R-to-S2}

\begin{document}
\title{Quantifying inner-outer interactions in non-canonical wall-bounded flows}
\author{Mogeng Li$^1$, Woutijn J. Baars$^1$, Ivan Marusic$^2$ and Nicholas Hutchins$^2$}
\affiliation{$^1$Faculty of Aerospace Engineering, Delft University of Technology, 2629 HS, The Netherlands\\
$^2$Department of Mechanical Engineering, University of Melbourne, Victoria 3010, Australia}

\date{\today}
\begin{abstract}
We investigate the underlying physics behind the change in amplitude modulation coefficient in non-canonical wall-bounded flows in the framework of the inner-outer interaction model (IOIM) (Baars \textit{et al.}, \textit{Phys. Rev. Fluids} \textbf{1} (5), 054406). The IOIM captures the amplitude modulation effect, and here we focus on extending the model to non-canonical flows. An analytical relationship between the amplitude modulation coefficient and IOIM parameters is derived, which is shown to capture the increasing trend of the amplitude modulation coefficient with an increasing Reynolds number in a smooth-wall dataset. This relationship is then applied to classify and interpret the non-canonical turbulent boundary layer results reported in previous works. We further present the case study of a turbulent boundary layer after a rough-to-smooth change. Both single-probe and two-probe hotwire measurements are performed to acquire streamwise velocity time series in the recovering flow on the downstream smooth wall. An increased coherence between the large-scale motions and the small-scale envelope in the near-wall region is attributed to the stronger footprints of the over-energetic large-scale motions in the outer layer, whereas the near-wall cycle and its amplitude sensitivity to the superposed structures are similar to that of a canonical smooth-wall flow. These results indicate that the rough-wall structures above the internal layer interact with the near-wall cycle in a similar manner as the increasingly energetic structures in a high-Reynolds number smooth-wall boundary layer.
\end{abstract}

\maketitle
\newpage

\section{Introduction}
In turbulent boundary layers, large coherent structures are found in the logarithmic region. They carry a high level of turbulent kinetic energy, and make a significant contribution to Reynolds stress production \citep{balakumar2007large}. These structures can be further classified as large-scale motions (LSMs) and very-large-scale motions (VLSMs). The former are associated with the vortex packets formed by aligned hairpin vortices \citep{Kim1999, Smits2011a}, and typically have a streamwise length of $\sim3\delta$ ($\delta$ is the boundary layer thickness), while the latter may be related to the merging of multiple LSMs \citep{Kim1999}, and can reach a streamwise extent up to $20\delta$ with spanwise meandering \citep{Hutchins2007,Hutchins2007a}. These structures are observed to leave a footprint in the near-wall region \citep{Hoyas2006, Hutchins2007, Hutchins2007a}. As the Reynolds number of a turbulent boundary layer increases, the separation between the outer-scaled motions and viscous-scaled near-wall cycle becomes more distinct, and the strength of these large-scale footprints also intensifies, resulting in the growth of the `inner-peak' magnitude in the broadband streamwise turbulence intensity \citep{Hutchins2007a, vincenti2013streamwise}. 

In addition to the direct superposition effect manifested as large-scale footprints, it has also been found that the near-wall small scales are modulated by the large scales in the outer layer \citep{bandyopadhyay1984coupling, Hutchins2007a, agostini2014influence}. In the near-wall region, the amplitude and frequency of small-scale fluctuations show a decrease when co-existing with large-scale low-speed regions, and vice versa in the case of large-scale high-speed regions. These observations were harnessed by the `two-scale' framework, where a local (near-wall) fine-mesh solution is coupled to the global coarse-mesh solution, in order to reduce computational costs at high Reynolds numbers \citep{he2018multiscale, chen2023two}. In terms of the modelling efforts, quasisteady quasihomogeneous theory \citep{zhang2016quasisteady, chernyshenko2021extension} provides an axiomatic description of the scale interaction in near-wall turbulence. The modulation effect was also quantified in recent works \citep{Mathis2009, Chung2010, guala2011interactions, ganapathisubramani2012amplitude, baars2015wavelet}, and a predictive model, termed the inner-outer interaction model (IOIM), which outputs representative turbulence statistics in the near-wall region based on an input signal in the logarithmic region, was developed by Marusic and coworkers \citep{Marusic2010a,Mathis2011} and by \citet{agostini2016predicting}. The former was later revised by \citet{baars2016spectral} using spectral linear stochastic estimation. The IOIM provides an opportunity to push the boundary of large-eddy simulations of wall-bounded flows to very high Reynolds numbers at affordable costs, thanks to its ability to provide representative real-time small-scale signals in the viscous-scaled near-wall region based only on large-scale outer-layer information \citep{inoue2012inner, cabrit2014towards, sidebottom2014modelling}.

Although similar amplitude modulation behaviours have been observed in turbulent boundary layers, channel, and pipe flows \citep{mathis2009comparison}, the existence of such scale interactions remains unexplored under non-canonical conditions. A sound understanding of how the inner- and outer-scale relationship is affected by these conditions is essential for generalising its application to a wider scope of flows. Enhanced amplitude modulation has been observed in various non-canonical flows, including boundary layers over rough walls \citep{monty2010high,squire2016inner,anderson2016amplitude, wu2019modelling, blackman2019assessment} or permeable surfaces \citep{kim2020experimental, khorasani2022turbulent}, and boundary layers with modified outer structures, such as energetic large-scale motions injected into the flow via freestream turbulence \citep{dogan2016interactions,dogan2017modelling}, upstream dynamic roughness \citep{duvvuri2015triadic} and synthetic large-scale signals generated by plasma actuators \citep{lozier2022experimental}, to name a few. It is not yet well understood how these seemingly different flow conditions all lead to a common increase in the amplitude modulation coefficient, and we aim to bridge this gap by establishing a physics-based quantitative relationship between the amplitude modulation coefficient and IOIM parameters.

\begin{figure}
    \centering
    \includegraphics[width=3in, clip]{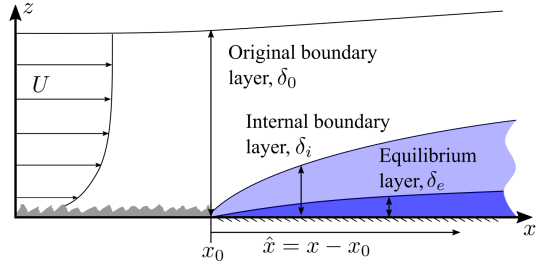}
    \caption{Schematic of a turbulent boundary layer over a rough-to-smooth change in surface condition. The roughness transition occurs at $x_0$, and $\hat{x} = x-x_0$ denotes the fetch downstream of the transition. Reproduced from \citet{MogengJFM2019}. }
    \label{fig:IBL_sketch}
\end{figure}

Furthermore, here we look into another scenario where the introduced large scales in the outer layer have an energy distribution across scales similar to that of a canonical boundary layer, and only the amplitude is intensified. This is achieved by a sudden rough-to-smooth surface transition occurring in the streamwise direction, as depicted in Fig. \ref{fig:IBL_sketch}. Upstream of the transition, a turbulent boundary layer develops on a rough wall with equivalent sand grain roughness height $k_s$. Here, $x$ is the streamwise direction, $x_0$ is the streamwise location of the surface transition and $\hat{x}\equiv x-x_0$ is the distance downstream of the transition. At $x = x_0$, the surface switches to a smooth wall, while the boundary layer continues to evolve and gradually adjusts to the new surface. The effect of the new surface condition is first felt in the near-wall region of the boundary layer and then gradually propagates to the interior of the flow \citep{Garratt1990}. The layer that separates the modified near-wall region from the unaffected oncoming flow farther away from the wall is generally referred to as the internal boundary layer (IBL) with a thickness denoted by $\delta_i$. For more details on the observation and modelling of the flow recovery, we refer to the works by \citet{Elliott1958}, \citet{antonia1972response}, \citet{Hanson2016}, \citet{rouhi2018}, and \citet{Mogengl2020, mogeng2022modelling}.

Turbulent boundary layers over a rough-to-smooth change in the streamwise direction offer a new perspective to further understand the physics of inner-outer interactions. When normalised by the friction velocity at the wall, LSMs and VLSMs above $\delta_i$ are similar to their smooth-wall counterparts \citep{Townsend1976, squire2016comparison}, but they are over-energised compared to the near-wall small scales. The former retain a memory of the upstream rough wall friction velocity whereas the latter scale on the much lower local smooth-wall friction velocity. How these structurally similar but more energised large-scale motions interact with the near-wall cycle will be investigated in this study.

The remainder of the paper is organised as follows: in Sec. \ref{sec:R_IOIM}, we first present a summary of the definition of the amplitude modulation coefficient and the IOIM framework, as well as a quantitative relation between the two. We then review the previous studies on the amplitude modulation in non-canonical flows in Sec. \ref{sec:discussions}. In Sec. \ref{sec:case_study}, we present a case study on the flow downstream of a rough-to-smooth change in the surface condition.

\section{Physical underpinning of amplitude modulation} \label{sec:R_IOIM}
In this section, we first briefly summarise the definition of the amplitude modulation coefficient $R$ and the IOIM formulation, and then provide a quantitative description of the relation between the two, supported by results computed from synthetic signals.

\subsection{Amplitude modulation coefficient and IOIM formulation}
\begin{figure}
\centering
\setlength{\unitlength}{1cm}
\begin{picture}(13.5,7.8)
\put(0,0){\includegraphics[scale = 0.9]{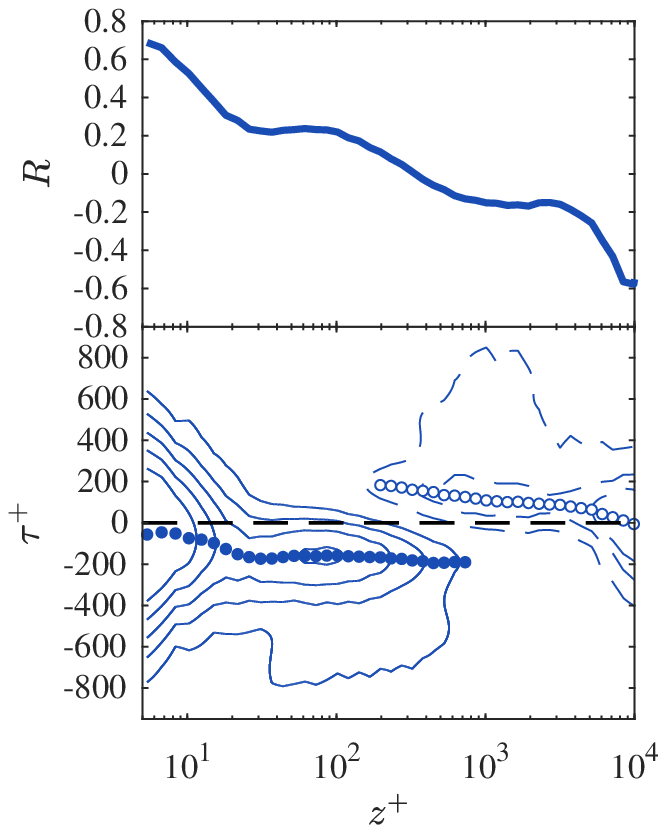}}
\put(7,0){\includegraphics[scale = 0.9]{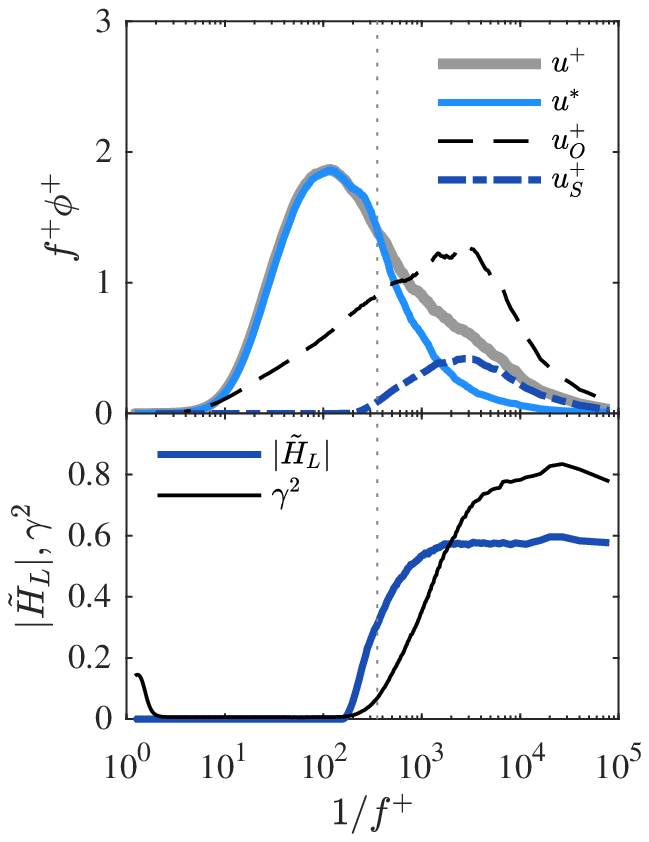}}
\put(0.1,7.5){(\textit{a})}
\put(7.3,7.5){(\textit{c})}
\put(0.1,4.4){(\textit{b})}
\put(7.3,3.9){(\textit{d})}
\put(5.75,2.78){\line(1,0){0.3}}
\put(6.05,2.78){\line(0,1){3.72}}
\put(6.05,6.5){\vector(-1,0){0.3}}
\put(10.5,7.7){$\lambda_{x}^+=7000$}
\end{picture}
\caption{(\textit{a}) Amplitude modulation coefficient $R(z^+)$ (Eq. \ref{eq:R}) of a smooth-wall turbulent boundary layer with $Re_{\tau} = 1.3\times10^5$ \citep{marusic2015evolution}. (\textit{b}) Isocontour of the time-shifted amplitude modulation coefficient $R_{\tau}(z^+,\tau^+)$ (Eq. \ref{eq:R_tau}) of the same boundary layer profile. The solid contour lines are from 0.1 to 0.5 with a step of 0.1, and the dashed contour lines are from $-0.3$ to $-0.1$ with a step of 0.1. The solid and filled circles mark the $\tau^+$ values where $R_{\tau}$ reaches its maximum and minimum at each wall-normal location, respectively. (\textit{c}) Premultiplied energy spectra of the measured velocity $u^+$, universal small scales $u^*$ and superposition $u_S^+$ at $z^+\approx10$, and $u_O^+$, the velocity measured by the outer probe at $z_O^+ = 469$. (\textit{d}) The gain of the linear kernel $|\widetilde{H}_L|$ and the linear coherence spectrum $\gamma^2(f^+)$ at $z^+\approx10$. The vertical dotted line in (\textit{c}) and (\textit{d}) marks the cut-off wavelength $\lambda_x^+=7000$ (with the mean velocity at the outer probe as the convective velocity) used by \citet{Mathis2011} to separate large and small scales.}
\label{fig:R_HL}
\end{figure}

The amplitude modulation coefficient $R$ is a commonly reported diagnostics, largely due to the fact that it provides a straightforward quantification of the degree of amplitude modulation within a single-point time series. An example of the $R$ profile is shown in Fig. \ref{fig:R_HL}(\textit{a}). It is defined as the correlation between the low-pass-filtered envelope of small-scale fluctuations and the large-scale fluctuations at the same location \citep{Mathis2009}:
\begin{equation}
R(z^+) = \frac{\left<E_L\left[u_d^+\left(z^+,t^+\right)\right] u_S^+\left(z^+,t^+\right)\right>}{\sqrt{\left<E_L^2\left[u_d^+\left(z^+,t^+\right)\right] \right>}\sqrt{\left<u_S^{+2}\left(z^+,t^+\right)\right>}}
\label{eq:R}
\end{equation}
Here, the superscript $(\cdot)^+$ indicates inner scaling with the local friction velocity as the velocity scale,  the angle brackets $\left<\cdot\right>$ denotes time average, $u$ represents the streamwise velocity, $u_S^+$ is the zero-mean large-scale superposition signal, which is usually obtained by low-pass filtering the time series with a threshold of $\lambda_x^+ = 7000$ and $u_d^+ \equiv u^+-u_S^+$ is the detrended signal. $E_L[\cdot]$ denotes a low-pass-filtered envelope of the signal
\begin{equation}
E_L\left[u_d^+\left(z^+,t^+\right)\right] = \mathcal{L}\left[\sqrt{u_d^{+2}\left(z^+,t^+\right)+\mathcal{H}^2\left[u_d^{+}\left(z^+,t^+\right)\right]}\right],
\label{eq:EL}
\end{equation} 
where $\mathcal{H}[\cdot]$ is the Hilbert transform, $\sqrt{u_d^{+2}\left(z^+,t^+\right)+\mathcal{H}^2\left[u_d^{+}\left(z^+,t^+\right)\right]}$ is the analytic signal and $\mathcal{L}[\cdot]$ denotes a low-pass filter.

The definition can be further generalised to a time-shifted amplitude modulation coefficient 
 \begin{equation}
R_{\tau}(z^+,\tau^+) = \frac{\left<E_L\left[u_d^+\left(z^+,t^+\right)\right] u_S^+\left(z^+,t^+-\tau^+\right)\right>}{\sqrt{\left<E_L^2\left[u_d^+\left(z^+,t^+\right)\right] \right>}\sqrt{\left<u_S^{+2}\left(z^+,t^+\right)\right>}}.
\label{eq:R_tau}
\end{equation}
By definition, $R(z^+) \equiv R_{\tau}(z^+,0)$ (see Fig. \ref{fig:R_HL}\textit{b}). The relative shift $\tau_a^+$ is the lag between the superposition imprint $u_S^+$ and the low-pass filtered envelope $E_L\left[u_d^+\right]$ such that $R_{\tau}$ reaches its maximum, i.e. $R_{\tau}(z^+,\tau_a^+) = \text{max}\left[R_{\tau}(z^+,\tau^+)\right]$, as marked by the solid circles in Fig. \ref{fig:R_HL}(\textit{b}). 

According to the IOIM \citep{baars2016spectral}, the statistical prediction of the fluctuating velocity $u_p^+$ can be constructed by considering a superposition effect of large-scale content with additively a universal signal $u^*$ that is subject to an amplitude modulation
\begin{equation}
u_p^+\left(z^+,t^+\right) = \underbrace{u^*\left(z^+,t^+\right) \left\{1+\Gamma\left(z^+\right)u_S^+\left(z^+,t^+-\tau_a^+\right) \right\}}_\text{amplitude modulation}+\underbrace{u_S^+\left(z^+,t^+\right)}_\text{superposition}.
\label{eq:IOIM}
\end{equation}
Here, $\Gamma$ is the amplitude sensitivity, and the large-scale imprints $u_S^+$ can be found from a given outer-layer large-scale signal $u_O^+\left(t^+\right)$ via
\begin{equation}
u_S^+\left(z^+,t^+\right) = \mathcal{F}^{-1}\left[\widetilde{H}_L\left(z^+,f^+\right)\mathcal{F}\left[u_O^+\left(t^+\right)\right]\right],
\label{eq:HL}
\end{equation}
where $\mathcal{F}[\cdot]$ and $\mathcal{F}^{-1}[\cdot]$ represent a Fourier transform and inverse Fourier transform, respectively, and $\widetilde{H}_L\left(f^+\right)$ is the linear transfer kernel incorporating the large-scale coherence between the near-wall and outer regions.

The gain of the linear kernel $|\widetilde{H}_L|$ is related to the linear coherence spectrum $\gamma^2(f^+)$ via
\begin{equation}
|\widetilde{H}_L(f^+)| = \sqrt{\gamma^2(f^+)\frac{\left<\left|\mathcal{F}\left[u^+\right]\right|^2\right>}{\left<\left|\mathcal{F}\left[u_O^+\right]\right|^2\right>}},
\end{equation}
where $\gamma^2(f^+)$ is given by \citet{bendat2011random}
\begin{equation}
\gamma^2(f^+) = \frac{\left|\mathcal{F}\left[u_O^+\right]\overline{\mathcal{F}\left[u^+\right]}\right|^2}{\left<\left|\mathcal{F}\left[u_O^+\right]\right|^2\right>\left<\left|\mathcal{F}\left[u^+\right]\right|^2\right>},
\end{equation}
with $\overline{\left(\cdot\right)}$ denoting the complex conjugate.

As shown in Figs. \ref{fig:R_HL}(\textit{c}) and (\textit{d}), only the large-scale energy of the outer velocity signal $u_O^+$ is retained in the superposition signal $u_S^+$, which contributes to the large-scale end of the energy spectrum of $u^+$. The small-scale end of the spectrum, on the other hand, is mainly from the universal small scale, $u^*$. The linear transfer kernel $|\widetilde{H}_L|$ enables a smooth roll-off of the coherence from large to small scales, and the scale separation here is around the commonly used cut-off threshold of $\lambda^+_x = 7000$.

\subsection{Quantitative relationship between $R$ and IOIM parameters}
In this subsection, we derive a quantitative relationship of the amplitude modulation coefficient by expressing $R$ in the framework of IOIM. Following \citet{duvvuri2015triadic}, we work with a modified expression of the amplitude modulation coefficient 
\begin{equation}
R_2(z^+) = \frac{\left<E_{2L}\left[u_d^+\left(z^+,t^+\right)\right] u_S^+\left(z^+,t^+\right)\right>}{\sqrt{\left<E_{2L}^2\left[u_d^+\left(z^+,t^+\right)\right] \right>}\sqrt{\left<u_S^{+2}\left(z^+,t^+\right)\right>}},
\label{eq:R2}
\end{equation}
where
\begin{equation}
E_{2L}\left[v\left(t^+\right)\right] = \mathcal{L}\left[v^{2}\left(t^+\right)+\mathcal{H}^2\left[v\left(t^+\right)\right]\right]
\label{eq:E2L}
\end{equation} 
for an arbitrary time series $v\left(t^+\right)$. The modified coefficient $R_2$ uses the square of the analytic signal to avoid the difficulty in dealing with the square root in Eq. (\ref{eq:EL}), leading to a simpler mathematical expression, and given that the amplitude modulation coefficient is a normalised measure, no significant difference is expected in the values of $R$ and $R_2$. This can be easily verified using experimental data.

By expressing the velocity signals $u_d^+$ and $u_S^+$ in a series of Fourier modes, and with some trigonometric manipulations, \citet{duvvuri2015triadic} showed that
\begin{equation}
R_2 = \frac{2\left<u_d^{+2}u_S^+\right>}{\sqrt{\left<E_{2L}^2\left[u_d^+\right] \right>}\sqrt{\left<u_S^{+2}\right>}}.
\label{eq:R2T}
\end{equation}

We substitute the detrended signal expressed using the notations of IOIM (i.e. $u_d^+ = u^*\left[1+\Gamma u_S^+\left(t^+-\tau_a^+\right)\right]$) into Eq. (\ref{eq:R2T}) and noting that $\left<u^{*2}u_S^{+}\right> = 0$ because $u^*$ and $u_S^+$ are uncorrelated by definition, $R_2$ can then be expressed as
\begin{equation}
R_2 = \frac{4\Gamma\left<u^{*2}\right>\left<u_S^+\left(t^+\right)u_S^+\left(t^+-\tau_a^+\right)\right>+2\Gamma^2\left<u^{*2}\right>\left<u_S^+\left(t^+\right)u_S^{+2}\left(t^+-\tau_a^+\right)\right>}{\sqrt{\left<\left\{E_{2L}\left[u^*\right]+2\Gamma\mathcal{L}\left[u^{*2}u_S^++\mathcal{H}\left[u^*\right]\mathcal{H}\left[u^*u_S^+\right]\right]+\Gamma^2E_{2L}\left[u^*u_S^+\right]\right\}^2 \right>}\sqrt{\left<u_S^{+2}\right>}}
\label{eq:R2_expand}
\end{equation}
Comparing the orders of $u^*$, $u_S^+$ and $\Gamma$ in the terms in the denominator of Eq. (\ref{eq:R2_expand}) and noting that both $\mathcal{L}[\cdot]$ and $\mathcal{H}[\cdot]$ are linear operators give rise to
\begin{subequations}
\begin{align}
E_{2L}\left[u^*\right] &\sim \left<u^{*2}\right>,\\
\Gamma\mathcal{L}\left[u^{*2}u_S^++\mathcal{H}\left[u^*\right]\mathcal{H}\left[u^*u_S^+\right]\right] &\sim \Gamma\left<u_S^{+2}\right>^{1/2}\left<u^{*2}\right>,\\
\Gamma^2E_{2L}\left[u^*u_S^+\right] &\sim \Gamma^2\left<u_S^{+2}\right>\left<u^{*2}\right>.
\end{align}
\end{subequations}
Typically, $\Gamma$ is a relatively small number ranging from $\mathcal{O}(0.01)$ to $\mathcal{O}(0.10)$ \citep{baars2016spectral}, and $\left<u_S^{+2}\right><\left<u^{*2}\right>$ especially in the near-wall region of $z^+\lesssim200$ which is the current focus. Therefore, terms containing higher orders of $\Gamma\left<u_S^{+2}\right>^{1/2}$ can be neglected, and Eq. (\ref{eq:R2_expand}) is then reduced to 
\begin{equation}
R_2 = \frac{4\Gamma\left<u^{*2}\right>\left<u_S^+\left(t^+\right)u_S^+\left(t^+-\tau_a^+\right)\right>}{\sqrt{\left<E^2_{2L}\left[u^*\right] \right>}\sqrt{\left<u_S^{+2}\right>}}
\label{eq:R2_reduce}
\end{equation}
We identify the following 3 parameters that can contribute to a change in $R_2$:
\begin{enumerate}
\item The relative shift $\tau_a^+$. 
\item The amplitude sensitivity $\Gamma$.
\item The amplitude of large-scale imprints $\left<u_S^{+2}\right>$.
\end{enumerate}
Interestingly, $R_2$ is not affected by changes in the amplitude of universal small scales $\left<u^{*2}\right>$. This is because both the numerator and the denominator in Eq. (\ref{eq:R2_expand}) (and Eq. (\ref{eq:R2_reduce}) as well) contain the same order of $\left<u^{*2}\right>$, which eventually cancel out. Note that changes in the energy distribution across scales in $u_S^+$ or $u^*$ have more complicated consequences: for $u_S^+$, it will affect the auto-correlation term $\left<u_S^+\left(t^+\right)u_S^+\left(t^+-\tau_a^+\right)\right>$ with a given $\tau_a^+$, and for $u^*$, it will affect how much energy remains in $E_{2L}\left[u^*\right]$ after a low-pass filter is applied. Therefore, we limit the quantitative analysis to changes in the amplitude of $u_S^+$ and $u^*$ fluctuations. 

In canonical smooth-wall turbulent boundary layers, the IOIM parameters ($u^*$, $\Gamma$, $\tau_a^+$, and $\widetilde{H}_L$) are Reynolds number invariant over the range of $Re_{\tau} \approx 7350 -13300$ tested in the calibrations \citep{baars2017reynolds}, and only a change in the large-scale imprint amplitude $\left<u_S^{+2}\right>$ was observed when varying the Reynolds number. However, $\tau^+_a$, $\Gamma$ and $\left<u_S^{+2}\right>$ can all be modified under non-canonical conditions, and it is important to understand how they individually contribute to the overall amplitude modulation coefficient $R$.

\subsubsection{Effect of the relative shift $\tau_a^+$}
We can conclude from Eq. (\ref{eq:R2_reduce}) that $R_2$ would increase with a decreasing $\left|\tau_a^+\right|$, as a result of reduced time lag, or improved alignment between modulation and superposition signals. Typically, the relative shift $\tau_a^+$ is much smaller than the time period of $u_S^+$ below the centre of the log region, where the predictive model is applied. Therefore, $u_S^+\left(t^+\right)$ and $u_S^+\left(t^+-\tau_a^+\right)$ are largely in phase. The auto-correlation term $\left<u_S^+\left(t^+\right)u_S^+\left(t^+-\tau_a^+\right)\right>$ is positive, and increases with a decreasing $\left|\tau_a^+\right|$. 

Fig. \ref{fig:R_Gamma_uS}(\textit{a}) shows $R$ (Eq. \ref{eq:R}) computed from the velocity signals constructed from $u^*$ and $u_S^+$ with a range of different $\tau_a^+$ values following (\ref{eq:IOIM}). The IOIM parameters and the baseline values of $\tau_{a0}^+$, $\Gamma_0$ and $\left<u_{S0}^{+2}\right>$ are taken from the dataset at $Re_{\tau} = 13300$ \citep{baars2016spectral}. The decreasing trend of $R$ with an increasing $\left|\tau_a^+\right|$ confirms the conclusion based on Eq. (\ref{eq:R2_reduce}).

\subsubsection{Effect of the amplitude sensitivity $\Gamma$}
According to Eq. (\ref{eq:R2_reduce}), $R_2$ will increase with an increasing $\Gamma$. This trend is corroborated by the increasing $R$ computed from the signals constructed following Eq. (\ref{eq:IOIM}) with an increasing $\Gamma/\Gamma_0$ (see Fig. \ref{fig:R_Gamma_uS}\textit{b}).

\subsubsection{Effect of the superposition intensity $\left<u_S^{+2}\right>$} \label{subsubsec:effect_u_S}
\begin{figure}
\centering
\setlength{\unitlength}{1cm}
\begin{picture}(14,4.8)
\put(0,0){\includegraphics[scale = 0.9]{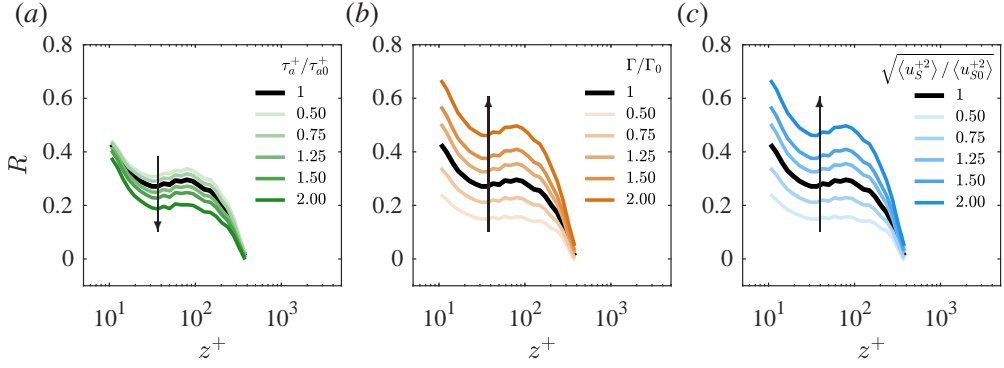}}
\put(0.1,4.5){(\textit{a})}
\put(4.5,4.5){(\textit{b})}
\put(8.9,4.5){(\textit{c})}
\put(2,2.7){\vector(0,-1){1}}
\put(6.4,1.7){\vector(0,1){1.8}}
\put(10.8,1.7){\vector(0,1){1.8}}
\end{picture}
\caption{Amplitude modulation coefficient $R$ of the velocity signal constructed with various (\textit{a}) $\tau_a^+$, (\textit{b}) $\Gamma$ and (\textit{c}) $\left<u_{S}^{+2}\right>$ values. The baseline parameters are denoted by $\tau_{a0}^+$, $\Gamma_0$ and $\left<u_{S0}^{+2}\right>$, respectively, and they are computed from the dataset at $Re_{\tau} = 13300$ in \citet{baars2016spectral}. The arrows in the panels indicate the direction of increasing the ratio between the varied parameter and the baseline.}
\label{fig:R_Gamma_uS}
\end{figure}

Based on Eq. (\ref{eq:R2_reduce}), $R_2$ will increase with an increasing $\left<u_S^{+2}\right>$. In fact, a closer examination of the Eqs. (\ref{eq:R2_reduce}) and (\ref{eq:R2_expand}) reveals that $\Gamma$ and $\left<u_S^{+2}\right>$ can be grouped into a single variable $\Gamma\left<u_S^{+2}\right>^{1/2}$. This is confirmed by the same trends of $R$ with $\Gamma$ and $\left<u_S^{+2}\right>^{1/2}$ in Figs. \ref{fig:R_Gamma_uS}(\textit{b}) and (\textit{c}), respectively. We note that the monotonic increase of $R_2$ with $\Gamma\left<u_S^{+2}\right>^{1/2}$ breaks down when $\Gamma\left<u_S^{+2}\right>^{1/2}$ becomes comparable with or larger than $\left<u^{*2}\right>^{1/2}$ (not shown in the figure), and the full expression of Eq. (\ref{eq:R2_expand}) can introduce non-monotonic dependence on $\Gamma\left<u_S^{+2}\right>^{1/2}$. However, given that $\Gamma$ is typically small, and $\left<u_S^{+2}\right>$ is much smaller than $\left<u^{*2}\right>$ in the near-wall region at the Reynolds number range investigated in the experimental dataset ($Re_{\tau}\lesssim2\times 10^4$), the non-monotonic is less likely to occur.

To summarise, we have shown analytically that the amplitude modulation coefficient will increase with (i) a reducing $\left|\tau_a^+\right|$, (ii) an increasing $\Gamma$, and (iii) an increasing $\left<u_S^{+2}\right>$, provided  $\Gamma \left<u_S^{+2}\right>^{1/2}\ll\left<u^{*2}\right>^{1/2}$.

\begin{figure}
\centering
\setlength{\unitlength}{1cm}
\begin{picture}(6.9,4.6)
\put(0,0){\includegraphics[scale = 0.95]{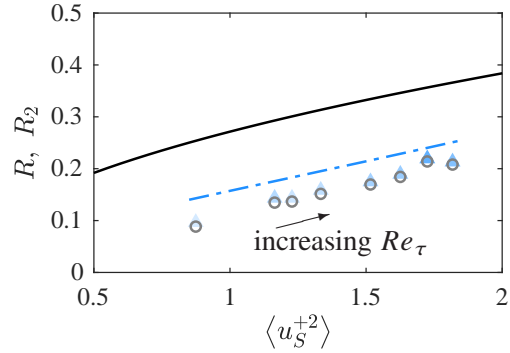}}
\put(3.5,1.6){\vector(4,1){0.7}}
\put(3.2,1.3){increasing $Re_{\tau}$}
\end{picture}
\caption{Amplitude modulation coefficients $R$ (grey circles) and $R_2$ (blue triangles) of a canonical smooth-wall turbulent boundary layer plotted against the corresponding superposition signal intensity $\left<u_S^{+2}\right>$ at $z^+ = 50$. From pale to dark blue, the shade of blue triangles indicates $Re_{\tau}$ increasing from 2800 to 13400. The solid black line is Eq. (\ref{eq:R2_reduce}) with a zero time lag ($\tau_a^+ = 0$), and the dot-dashed blue line is Eq. (\ref{eq:R2_reduce}) with the auto-correlation term $\left<u_S^+\left(t^+\right)u_S^+\left(t^+-\tau_a^+\right)\right>$ estimated from the experimental data.}
\label{fig:R_uS_Sdata}
\end{figure}

\subsection{Verification with a smooth-wall turbulent boundary layer dataset}
The analysis above is verified using an experimental smooth-wall turbulent boundary layer dataset with $Re_{\tau}$ ranging from 2800 to 13400 \citep{marusic2015evolution}. In this series of canonical smooth-wall boundary layer profiles, both $\tau_a^+$ and $\Gamma$ are expected to remain constant, and only $\left<u_S^{+2}\right>$ increases with an increasing $Re_{\tau}$ as a consequence of more energetic large-scale motions in the outer layer, providing an ideal test ground to examine the dependence of $R$ on $\left<u_S^{+2}\right>$. Fig. \ref{fig:R_uS_Sdata} shows the amplitude modulation coefficients $R$ and $R_2$ computed directly from the velocity time series. The large-scale imprint intensity $\left<u_S^{+2}\right>$ is approximated by $\left<u^{+2}\right>-\left<u^{*2}\right>$, where $\left<u^{*2}\right>$ is the intensity of the universal small-scale signal from the calibration data of \citet{baars2016spectral}. The large-scale imprint intensity $\left<u_S^{+2}\right>$ increases with an increasing $Re_{\tau}$ as expected. The two coefficients $R$ and $R_2$ have very similar values, and they exhibit an increasing trend with $\left<u_S^{+2}\right>$, which is well captured by Eq. (\ref{eq:R2_reduce}). The auto-correlation term $\left.\left<u_S^+\left(t^+\right)u_S^+\left(t^+-\tau_a^+\right)\right>\right/\left<u_S^{+2}\right>$ in Eq. (\ref{eq:R2_reduce}) is smaller than 1 for a finite time shift $\tau_a^+$, but the exact value need to be determined from the time series of $u_S^+$. Assuming a perfect auto-correlation (which is essentially Eq. (\ref{eq:R2_reduce}) with a zero time lag, $\tau_a^+ = 0$) results in an overestimated $R_2$ as shown by the black line. The auto-correlation term at $z^+ \approx 50$ computed from the experimental data increases slightly from 0.56 to 0.69 from the lowest to the highest $Re_{\tau}$ measurements, because the relative phase shift $\left|\tau_a^+\right|U^+/\lambda_{S}^+$ reduces with increasing $Re_{\tau}$, with $\left|\tau_a^+\right|$ remaining constant and $\lambda_{S}^+$, the most energetic wavelengths of $u_S^+$, growing with the boundary layer thickness $\delta^+$ ($\equiv Re_{\tau}$). After taking this into account, the prediction of Eq. (\ref{eq:R2_reduce}) (dot-dashed blue line) captures the trend of data points very well. In summary, the increase in $R$ and $R_2$ with increasing $Re_{\tau}$ for a canonical smooth-wall turbulent boundary layer is primarily originated from the growing $\left<u_S^{+2}\right>$, and the slight increase in the auto-correlation term also makes a small contribution to the growth.

\section{Revisiting scale interactions in previous studies} \label{sec:discussions}
In this section, we present a summary of data from the literature and the current work, focusing on how the IOIM parameters change in various flow types and with increasing $Re_{\tau}$, and discuss their commonalities and differences in the scale interaction mechanism.

Based on whether the near-wall cycle is modified from that of an impermeable smooth wall, various flow types collected in Table \ref{tab:2} can be further classified as `top-down' and `bottom-up' categories, which indicate whether the deviation from a canonical flow is introduced in the outer layer (`top-down') or near to the wall (`bottom-up'). We note that although \citet{duvvuri2015triadic} used a dynamic roughness element upstream to generate the synthetic signal, the wall remains smooth at the location where the scale interaction is quantified, and therefore, this study is also classified as a `top-down' type. 

Notably, a higher amplitude modulation coefficient $R$ is found in all non-canonical flows collected in the table. A closer examination of the IOIM parameters would reveal that this increase in $R$ is caused by different mechanisms in each category. For rough-wall flows, although the strength of $\left<u_O^{+2}\right>$ is similar to that of a smooth-wall boundary layer, the coherence between the inner and outer layer is reduced, presumably due to the interruption of the shedded vortices in the roughness sublayer. The higher $R$ value is contributed by the increased $\Gamma$, which can be explained as follows \citep{squire2016inner}. In smooth-wall flows, small-scale fluctuations scale with viscous units. Considering that the large-scale variation of friction velocity $u_{\tau}$ is quasi-steady from the viewpoint of small scales,  the small-scale velocity fluctuations will then scale with $u_{\tau}$. By definition, $u_{\tau}\propto\sqrt{\tau_w}$, where $\tau_w$ is the instantaneous wall-shear stress and it is caused by and proportional to the large-scale velocity signature $u_S$ \citep{Mathis2013,zhang2016quasisteady}. Thus, the small-scale velocity fluctuations are proportional to $\sqrt{u_S}$. In fully-rough flows, however, the small-scale fluctuations are the wake vortices originated from roughness elements, the intensity of which is proportional to $u_S$. Therefore, the modulation sensitivity in rough-wall flows is expected to be stronger than that in smooth-wall flows. Following a similar argument, the near-wall region of a permeable substrate is dominated by upwelling/downwelling associated with the large-scale streamwise motions ($\propto u_S$) \citep{kim2020experimental,khorasani2022turbulent}, leading to an increase in $R$ as well.

Regarding the `top-down' flow category, more energetic large scales can be introduced in the outer layer by increasing $Re_{\tau}$ in a canonical smooth-wall boundary layer, through FST, or an upstream dynamic roughness element. Another example in this category, to be discussed in Sec. \ref{sec:case_study}, is that of a rough-to-smooth change in the wall condition, where the introduced large scales are the remnant of upstream rough-wall large-scale structures. The near-wall cycle (represented by the intensity of $u^*$) is little modified in all cases. The coherence between large scales in the inner and outer regions, which manifests in $|\widetilde{H}_L|$, either remains the same (canonical smooth wall with increasing $Re_{\tau}$), or increases (FST), and they all lead to stronger footprints $\left<u_S^{+2}\right>$ close to the wall. The stronger $\left<u_S^{+2}\right>$ is primarily responsible for the increased $R$ in the `top-down' flow cases.
\begin{turnpage}
\begin{table}
\begin{center}
\begin{ruledtabular}
\begin{tabular*}{\linewidth}{rllllllll}
\vspace{3mm}
&Flow type                           & Reference & $\left<u_O^{+2}\right>$   & $\left<u_S^{+2}\right>$  & $|\widetilde{H}_L|$ & $\left<u^{*2}\right>$                             & $\Gamma$  & $R$ (near wall) \\
\vspace{3mm}
\ldelim\{{6}{13mm}[top-down]&R-to-S change              &    present study       & higher  & higher & higher                            & similar                         & similar & higher          \\
\vspace{3mm}
&\begin{tabular}[c]{@{}l@{}}Smooth wall,\\ increasing $Re_{\tau}$ \end{tabular}&  \begin{tabular}[c]{@{}l@{}}\citet{Mathis2009, Mathis2011}\\ \citet{baars2016spectral}  \end{tabular}      & higher  & higher & similar                         & similar                         & similar & higher          \\
\vspace{3mm}
&FST                                 &   \citet{dogan2016interactions, dogan2017modelling}        & higher  & higher & higher                            & similar                         & lower     & higher          \\
\vspace{3mm}
&\begin{tabular}[c]{@{}l@{}}Single frequency\\ large-scale input \end{tabular}            &    \citet{duvvuri2015triadic}       & higher  &     higher   &          ?                         &        similar                           &   ?        &   ?              \\
\vspace{3mm}
\ldelim\{{5}{15mm}[bottom-up]&Rough wall                          &  \begin{tabular}[c]{@{}l@{}l@{}}\citet{squire2016inner}\\ \citet{anderson2016amplitude}\\  \citet{pathikonda2017inner} \\ \citet{blackman2019assessment}\end{tabular}        & similar & lower  & lower                             &  \begin{tabular}[c]{@{}l@{}l@{}}dependent on\\ roughness \\morphorlogy\end{tabular}& higher    & higher          \\
\vspace{3mm}
&Permeable wall                      &  \begin{tabular}[c]{@{}l@{}} \citet{efstathiou2018mean} \\ \citet{kim2020experimental}  \end{tabular}    &   similar/lower      &   ?     &           ?                        &          \begin{tabular}[c]{@{}l@{}l@{}}dependent on\\ substrate \\morphorlogy \end{tabular}    & ?   &     higher           
\end{tabular*}
\caption{Summary of IOIM parameters in various flow types. Quantities that cannot be inferred from the reported results are marked by `?'. Changes in the parameters are relative to a smooth-wall turbulent boundary layer at comparable Reynolds numbers. In the category of smooth-wall boundary layers with increasing Reynolds numbers, the comparison is made with regard to a lower Reynolds number. Results of the current study will be presented in details in Sec. \ref{sec:case_study}.}
\label{tab:2}
\end{ruledtabular}
\end{center}
\end{table}
\end{turnpage}

\begin{table}
\begin{center}
\def~{\hphantom{0}}
\begin{ruledtabular}
\begin{tabular}{lcccccccccc}
Case  & Colour & $\hat{x}/\delta_0$ & $Re_{\tau}$ & $3.9\sqrt{Re_{\tau}}$ & $U_{\infty}$ & $U_{\tau}$ & $z_O^+$ & $\delta_i^+$ & $l_I^+$ & $l_O^+$\\
 &  &  &  &  & ($\mathrm{ms}^{-1}$) & ($\mathrm{ms}^{-1}$) &  &  &  & \\
Smooth & black & - & 7350 & 334 & 10.0 & 0.34 & 441 & - & 22 & 22\\
\RtoSLow & green & 2.3 & 7200 & 330 & 22.5 & 0.69 & 331 & 890 & 21 & 21\\
\RtoSHigh  & magenta & 2.3 & 9600 & 382 & 31.0 & 0.92 & 388 & 1180 & 19 & 31\\
\end{tabular}
\caption{Experimental parameters of the two-point hotwire measurements, where the outer probe remains at a fixed location $z_O^+$ and the inner probe is traversed between $[0, z_O^+]$. $l_I^+$ and $l_O^+$ are the viscous-scaled hotwire filament length of the inner and outer probes, respectively.}
\label{tab:1}
\end{ruledtabular}
\end{center}
\end{table}

\section{Case study: turbulent boundary layer following a step change in surface roughness} \label{sec:case_study}
In this section, we present the new experimental results of a turbulent boundary layer downstream of a rough-to-smooth change in the surface conditions, which is another non-canonical flow configuration in the `top-down' category. We will briefly present the experimental conditions in Sec. \ref{sec:exp_setup}, and then explore the detailed modulation behaviours using both single-probe and two-probe results in Secs. \ref{sec:single_probe}--\ref{sec:two_probe}. The results are briefly discussed in Sec. \ref{subsec:case_discussion}.

\subsection{Experimental setup} \label{sec:exp_setup}
Simultaneous two-probe hotwire anemometry measurements are performed in the High Reynolds Number Boundary Layer Wind Tunnel (HRNBLWT) at the University of Melbourne. An upstream portion of the 27 m working section floor is covered by P24 sandpaper (SP40F, Awuko Abrasives) from the inlet to $x_0=7.2$ m, while the remaining length is a smooth aluminium surface. The peak-to-trough roughness height is $k_p\approx 1.2$ mm, and the equivalent sand grain roughness is $k_s\approx2.43$ mm. A nominal zero-pressure gradient is achieved by adjusting the bleeding slots on the tunnel roof. More details of the facility can be found in \citep{Mogengl2020} and \citep{Kulandaivelu2012}.

Two-probe hotwire measurements are performed at two freestream velocities, 22.5 m s$^{-1}$ and 31.0 m s$^{-1}$, and these two cases are named as \RtoSLow{} and \RtoSHigh{}, respectively. The flow conditions of these cases correspond to the single-point dataset of cases Re10ks16 and Re14ks22 in \citep{Mogengl2020}, and the local friction velocity $U_{\tau}$ of the current cases is interpolated from the skin-friction versus $\hat{x}$ trajectory in the same study, which was measured directly at the wall using oil-film interferometry. The procedures of the two-probe hotwire anemometry measurements are similar to that described by \citet{Mathis2011} and \citet{baars2016spectral}. The outer probe is fixed at $z_O^+=3.9\sqrt{Re_{\tau}}$, which is the geometric centre of the logarithmic region where the large-scale motions are highly active. The inner probe is traversed from the wall to below the outer probe with approximately 20 logarithmically spaced points in between. Both probes are conventional single-wire hotwire probes with a Wollaston wire etched to expose the sensing element. The length-to-diameter ratio of the exposed filament is $l/d\geq200$ \citep{ligrani1987spatial}.  For the case \RtoSLow{}, both wires have a diameter of $d=2.5 \upmu$m, leading to a viscous-scaled filament length of $l_I^+,l_O^+\approx 21$. For the case \RtoSHigh{}, $d=1.5\upmu$m wire is selected for the inner probe to maintain a similar spatial resolution with $l_I^+\approx 19$ with an increased freestream (and friction) velocity, while the outer probe filament diameter remains $d=2.5 \upmu$m because only the large-scale signal at this location is of the interest. Both probes are conventional single-wire hotwire probes operated by an in-house Melbourne University Constant Temperature Anemometer (MUCTA). The hotwire sampling time $T_s$ is more than 20000 boundary layer turn-over time ($\delta_{99}/U_{\infty}$) to achieve a good convergence of the statistics. A two-probe smooth-wall dataset from \citet{Mathis2011} and \citet{baars2016spectral} is also included for comparison in this study. Parameters of the two-point measurements are summarised in Table \ref{tab:1}.

\begin{figure}
\centering
\setlength{\unitlength}{1cm}
\begin{picture}(13,8)
\put(0,0){\includegraphics[scale = 0.95]{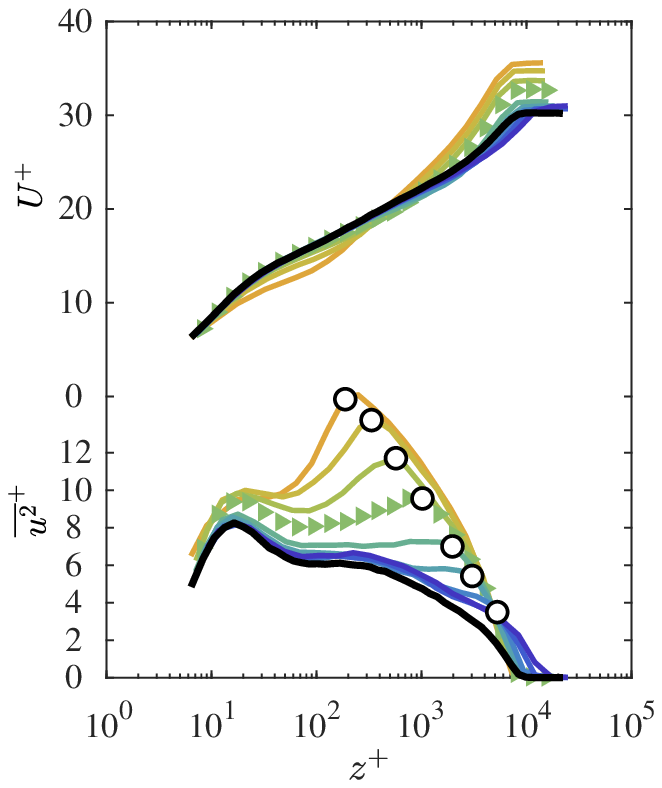}}
\put(6.5,0){\includegraphics[scale = 0.95]{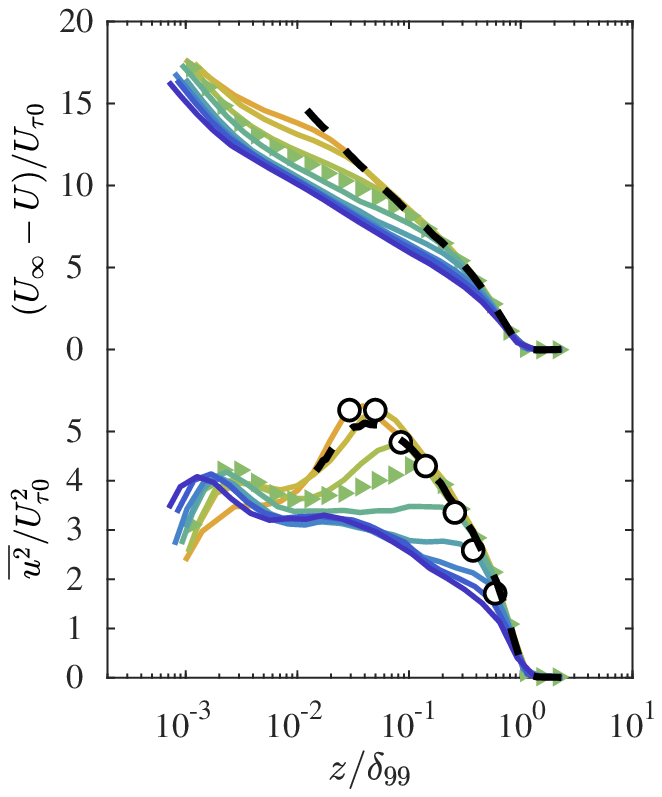}}
\put(0.1,7.7){(\textit{a})}
\put(6.6,7.7){(\textit{b})}
\end{picture}
\caption{Profiles of mean streamwise velocity and turbulence intensity corresponding to the flow conditions of \RtoSLow{}. (\textit{a}) is inner scaled using the local smooth-wall $U_{\tau}$, while (\textit{b}) is outer scaled, but using $U_{\tau0}$, the friction velocity measured just upstream of the rough-to-smooth transition. Line colours indicate the fetch, from orange to green to blue corresponds to $\hat{x}/\delta_0 = 0.2$, 0.5, 0.9, 1.9, 4.2, 7.4, 14.8, 29.2 and 53.1. The solid black line is a smooth-wall reference with $Re_{\tau} = 1.0\times10^4$ acquired in the same facility and normalised using the corresponding smooth-wall friction velocity \citep{marusic2015evolution}, and dashed black line is a rough-wall reference acquired just upstream of the rough-to-smooth change and normalised by the rough-wall friction velocity. The white circles represent the edge of the IBL, determined from the variance profile \citep{Mogengl2020}. The profile highlighted by triangular symbols is close to the streamwise location where the two-probe measurements in this study are performed.}
\label{fig:U_uvar}
\end{figure}

\subsection{Single-probe results} \label{sec:single_probe}

We first present the evolution of the mean velocity and turbulence statistics downstream of the rough-to-smooth change in Fig. \ref{fig:U_uvar}. Generally speaking, immediately after the rough-to-smooth change, the rough-wall turbulent boundary layer starts adapting to the new surface conditions first at the wall, and as the modified region enlarges, to the interior of the flow. The skin-friction coefficient experiences an undershoot before gradually increasing to the smooth-wall value. Fig. \ref{fig:U_uvar}(\textit{a}) shows the inner-scaled mean velocity profiles, where the velocity scale is selected as the friction velocity $U_{\tau}$ measured locally with oil-film interferometry over the smooth surface. Note that in this paper, the superscript $(\cdot)^+$ indicates inner scaling with the local friction velocity as the velocity scale. Fig. \ref{fig:U_uvar}(\textit{b}) shows the outer-scaled profiles, where the velocity scale is chosen as the friction velocity $U_{\tau 0}$ measured on the rough wall just upstream of the rough-to-smooth transition at $\hat{x}\rightarrow 0^{-}$. The inner-scaled mean velocity profiles of Fig. \ref{fig:U_uvar}(\textit{a}) collapse with the smooth-wall reference first close to the wall, while the outer-scaled profiles in Fig. \ref{fig:U_uvar}(\textit{b}) agree well with the rough-wall reference above the IBL (marked by the open circles). In the turbulence intensity profiles, a strong `outer-peak' manifests at the IBL, which is a result of the remaining energetic rough-wall structures. These structures leave a strong footprint in the near-wall region, as evidenced by the increased inner-peak magnitude in wall units shown in Fig. \ref{fig:U_uvar}(\textit{a}). The case \RtoSHigh{} has a higher $Re_{\tau0}$ and $k_{s0}^+$ compared to the case \RtoSLow{}, thus, stronger large-scale motions (which scale on $U_{\tau0}$) above the internal boundary layer are expected in the former. 

\begin{figure}
\centering
\includegraphics[scale = 0.95]{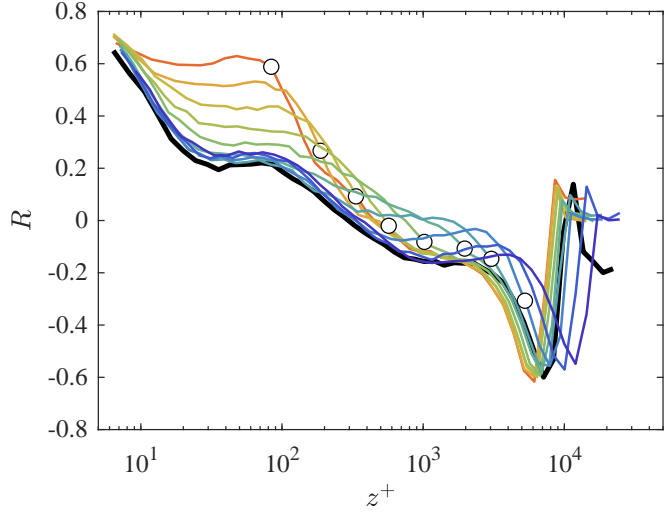}
\caption{Amplitude modulation coefficient $R$ at various streamwise locations downstream of a rough-to-smooth change. Line colours indicate the fetch, from red to blue corresponds to $\hat{x}/\delta_0 = 0.08,$ 0.2, 0.5, 0.9, 1.9, 4.2, 7.4, 14.8, 29.2 and 53.1. The white circles represent the edge of the IBL. The solid black line is a smooth-wall reference with $Re_{\tau} = 1.0\times10^4$ \citep{marusic2015evolution}.}
\label{fig:R}
\end{figure}

\begin{figure}
\centering
\setlength{\unitlength}{1cm}
\begin{picture}(13,5.5)
\put(0,0){\includegraphics[scale = 0.9]{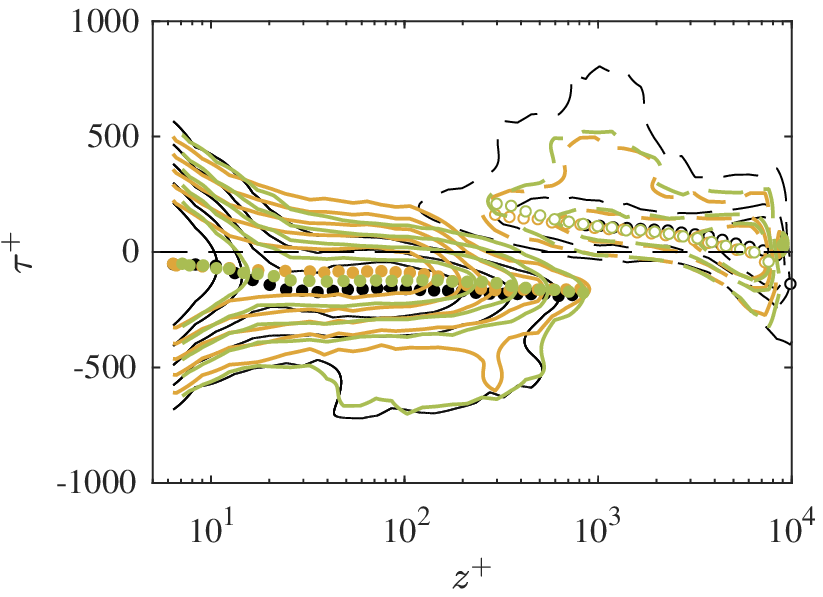}}
\put(8,0){\includegraphics[scale = 0.9]{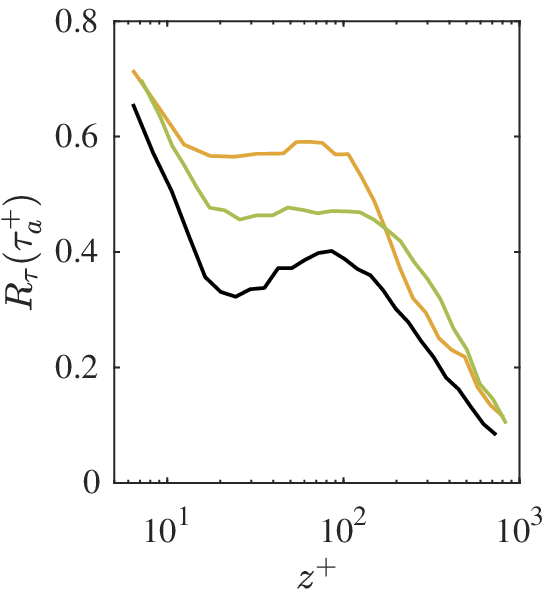}}
\put(0.1,5.2){(\textit{a})}
\put(7.9,5.2){(\textit{b})}
\end{picture}
\caption{(\textit{a}) Isocontours of $R_{\tau}$ at $\hat{x}/\delta_0 = 0.2$ (orange) and 0.9 (green) and the smooth-wall reference (black). The solid contour lines are from 0.1 to 0.5 with a step of 0.1, and the dashed contour lines are from $-0.3$ to $-0.1$ with a step of 0.1. The solid and empty circles mark the $\tau^+$ values where $R_{\tau}$ reaches its maximum and minimum at each wall-normal location, respectively. (\textit{b}) Maximum $R_{\tau}$ values, which are essentially $R_{\tau}$ at the locations marked by solid circles in (\textit{a}).}
\label{fig:R_2d}
\end{figure}

\begin{figure}
\centering
\setlength{\unitlength}{1cm}
\begin{picture}(13,5)
\put(0,0){\includegraphics[scale = 0.9]{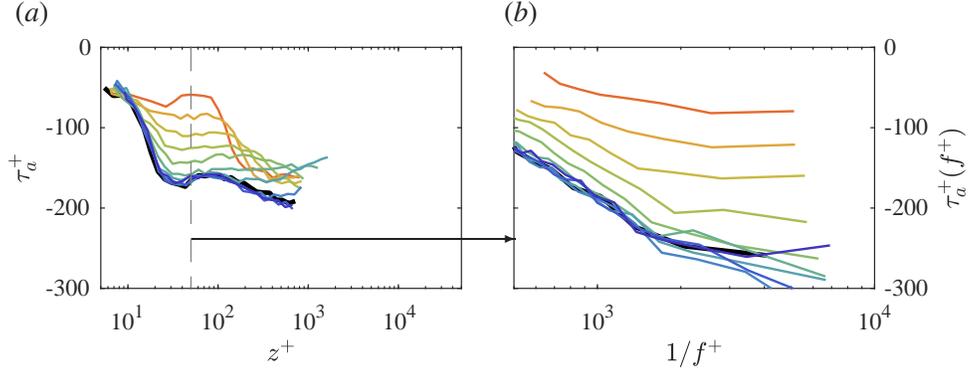}}
\put(0.1,4.6){(\textit{a})}
\put(6.6,4.6){(\textit{b})}
\put(2.45,1.7){\vector(1,0){4.3}}
\end{picture}
\caption{(\textit{a}) Time shift required for $R_{\tau}$ to reach its maximum. (\textit{b}) Argument of the complex co-spectrum of $u_S^+$ and $E_L[u_d^+]$ at $z^+ = 50$ (marked in (\textit{a}) by the vertical dashed line), premultiplied by the time period $1/f^+$. Legends are the same as in Fig. \ref{fig:R}. The solid black line is a smooth-wall reference with $Re_{\tau} = 1.3\times10^4$ \citep{marusic2015evolution}.}
\label{fig:tau_a_plus}
\end{figure}
The amplitude modulation coefficient $R$ (Eq. \ref{eq:R}) at a range of downstream locations from $\hat{x}/\delta_0 = 0.08$ to 53.1 is shown in Fig. \ref{fig:R}. The cut-off wavelength of the large-scale filter $\mathcal{L[\cdot]}$ is computed using the local friction velocity. Only the results of \RtoSLow{} are presented here for brevity, while the \RtoSHigh{} case shows similar behaviour. At small $\hat{x}/\delta_0$, a high $R$ is observed in the near-wall region. The coefficient $R$ decreases with an increasing $\hat{x}/\delta_0$, and beyond $\hat{x}/\delta_0>20$, it becomes very similar to that of a smooth-wall boundary layer.

Isocontours of $R_{\tau}$ (Eq. \ref{eq:R_tau}) at $\hat{x}/\delta_0 = 0.2$ (orange) and 0.9 (green) are shown in Fig. \ref{fig:R_2d}. In addition to the difference in the magnitude, at small fetches, the contours of positive correlations also shifts to the positive $\tau^+$ direction. In other words, close to the rough-to-smooth change, there is a smaller time lag between the envelope of the small-scale fluctuations and large-scale motions. The optimal positive correlation achieved at the time shift $\tau_a^+$ is shown in Fig. \ref{fig:R_2d}(\textit{b}). Similar to the zero-time-shift $R$ (shown in Fig. \ref{fig:R}), the maximum $R_{\tau}$ values are also higher at smaller fetches, implying that the high magnitudes of $R$ observed in Fig. \ref{fig:R} is more than the consequence of a smaller lag between the envelope and large-scale motions. 

A comparison of $\tau_a^+$ (time shift required for the optimum positive correlation) can be found in Fig. \ref{fig:tau_a_plus}(\textit{a}), where $\tau_a^+$ becomes more negative with an increasing fetch. Fig. \ref{fig:tau_a_plus}(\textit{b}) shows the argument of the complex co-spectrum $\phi$ at $z^+ = 50$, premultiplied by the time period $1/f^+$, following \citet{jacobi2013phase, jacobi2017phase} and \citet{deshpande2022relationship}. The co-spectrum is defined as $\phi\equiv\left<\mathcal{F}\left[u_S^+\right]\overline{\mathcal{F}\left[E_L[u_d^+]\right]}\right>$, and it can be viewed as the spectral equivalent of the amplitude modulation coefficient $R$. The premultiplied argument $\tau_a^+(f^+)\equiv\text{arg}(\phi)/(2\pi f^+)$ is essentially the time shift between $u_S^+$ and $E_L[u_d^+]$ in each Fourier mode, and it is reasonable that the overall time shift $\tau_a^+$ falls in the same range as $\tau_a^+(f^+)$ at each corresponding measurement location. In addition, the increase of the time lag with increasing fetch is also apparent here, confirming the trend of $\tau_a^+$ in Fig. \ref{fig:tau_a_plus}(\textit{a}). Further, the absolute time lag $\left|\tau_a^+(f^+)\right|$ is smaller at higher frequencies, meaning that the amplitude of small scales are more in-phase with the higher-frequency modes of the large scales. The increased maximum $R_{\tau}$ values at smaller fetches can partially be explained by the smaller scatter of $\tau_a^+(f^+)$ across a range of frequencies, because a single time shift $\tau_a^+$ can better align all Fourier modes in the $u_S^+$ and $E_L[u_d^+]$ signals in these cases.

The increase in $R$ compared to the smooth-wall reference has been previously observed in rough-wall flows and attributed to the stronger correlation between the amplitude of the small-scale turbulence associated with the roughness elements and the large-scale motions \citep{squire2016inner}. Compared to the most downstream location, $R$ in the logarithmic region is still noticeably higher at $\hat{x}/\delta_0 = 0.9$, which is equivalent to $\hat{x}/k_p\approx110$, a fetch where we might expect a large portion of the small-scale motions directly generated from the flow interaction with the roughness elements to diminish. However, small scales may form through the shear between the surviving rough-wall structures, and exhibit a stronger amplitude modulation effect with those structures from which they originate. 

In summary, based on the single-point measurements of the rough-to-smooth cases, the increase in $R$ is contributed by both reduced $|\tau_a^+|$ and increased $\left<u_S^{+2}\right>$, which are two out of the three factors identified in the analysis in Sec. \ref{sec:R_IOIM}. Further two-probe measurements are required to quantify the effect of the modulation sensitivity $\Gamma$.

\begin{figure}
\centering
\includegraphics[scale = 0.8]{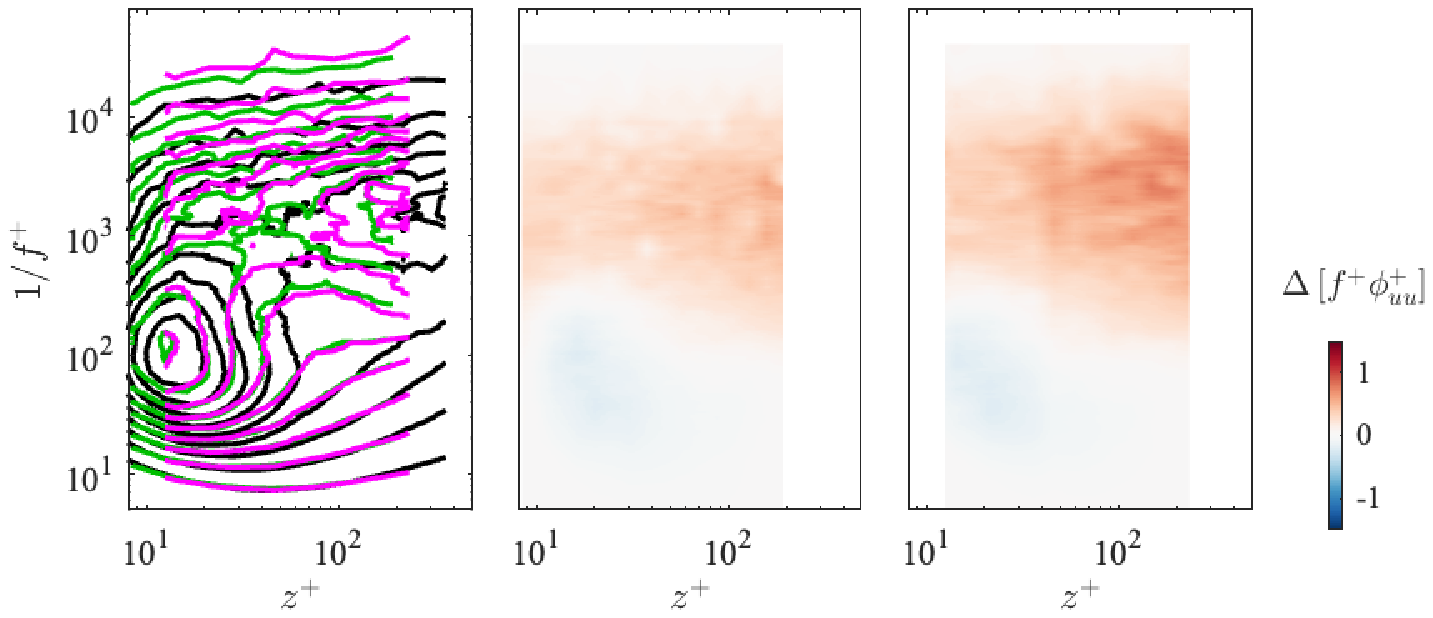}
\put(-310,133){(\textit{a})}
\put(-220,133){(\textit{b})}
\put(-130,133){(\textit{c})}\\
\includegraphics[scale = 0.8]{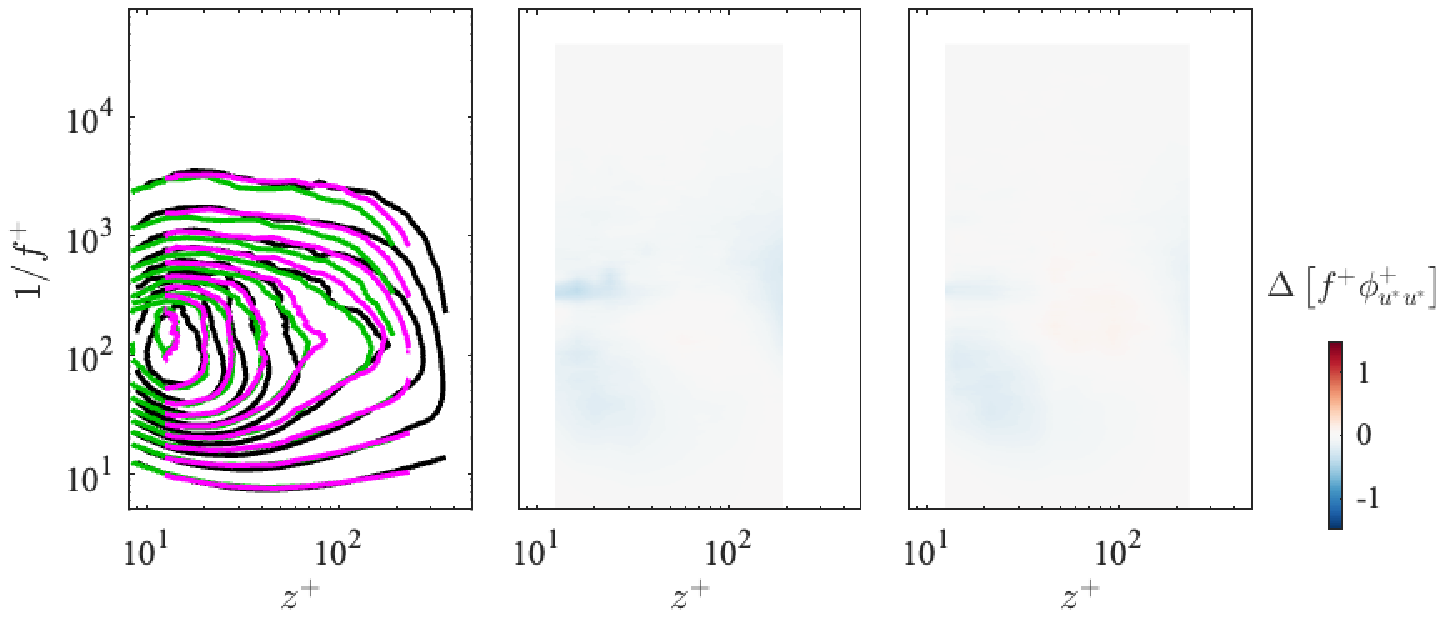}
\put(-310,133){(\textit{d})}
\put(-220,133){(\textit{e})}
\put(-130,133){(\textit{f})}\\
\caption{Premultiplied spectra of (\textit{a}) $u^+$ and (\textit{d}) $u^*$. The contour levels in both plots are from 0.2 to 2 with a spacing of 0.2. Line colours black, green and magenta represent the smooth, \RtoSLow{} and \RtoSHigh{} cases, respectively. (\textit{b}) and (\textit{c}) are the difference in $f^+\phi^+_{uu}$ between the \RtoSLow{} and \RtoSHigh{} cases and the smooth-wall reference, and (\textit{e}) and (\textit{f}) are the corresponding difference in $ f^+\phi^+_{u^*u^*}$.}
\label{fig:spectra}
\end{figure}
\subsection{Two-probe results} \label{sec:two_probe}
To further understand the origin of the enhanced modulation effect observed in Sec. \ref{sec:single_probe}, IOIM calibration following the procedure detailed in \citet{baars2016spectral} is performed on the three cases listed in Table \ref{tab:1}. In the results below, we will be using black for the smooth-wall reference, green and magenta for the \RtoSLow{} and \RtoSHigh{} cases, respectively.

Premultiplied energy spectra of the measured velocity fluctuation and the universal small-scale signal are shown in Figs. \ref{fig:spectra}(\textit{a}) and (\textit{d}), respectively. Figs. \ref{fig:spectra}(\textit{b}) and (\textit{c}) are the difference between the rough-to-smooth and smooth-wall reference. A band with excess energy at $1/f^+\approx2000$ is interpreted as the large-scale footprints in the near-wall region. The universal small scale spectra of the three cases (Fig. \ref{fig:spectra}\textit{d}) are very similar, indicating a re-establishment of the near-wall cycle after the rough-to-smooth change.

\begin{figure}
\centering
\includegraphics[scale = 0.85]{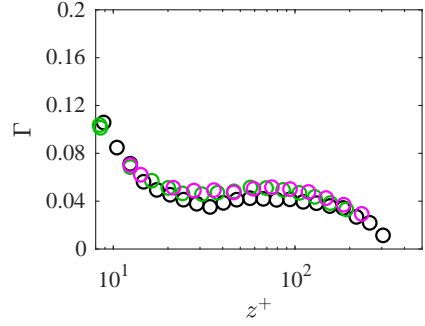}
\caption{Amplitude sensitivity $\Gamma(z^+)$. Line colours black, green and magenta represent the smooth, \RtoSLow{} and \RtoSHigh{} cases, respectively.}
\label{fig:Gammavec}
\end{figure}

\begin{figure}
\centering
\includegraphics[scale = 0.85]{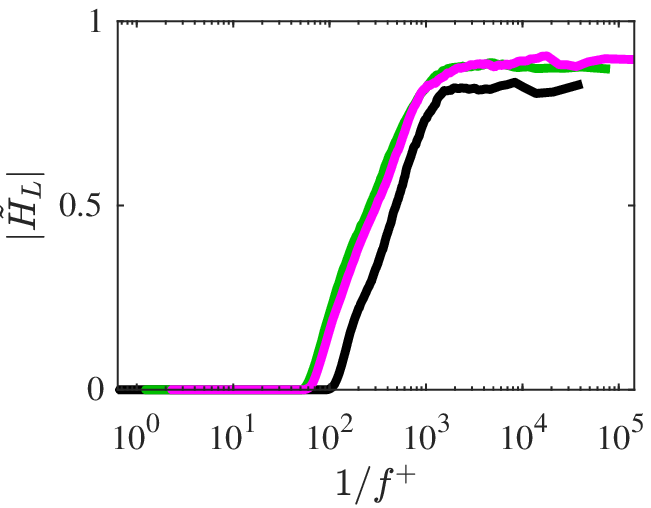}
\hspace{5mm}
\includegraphics[scale = 0.85]{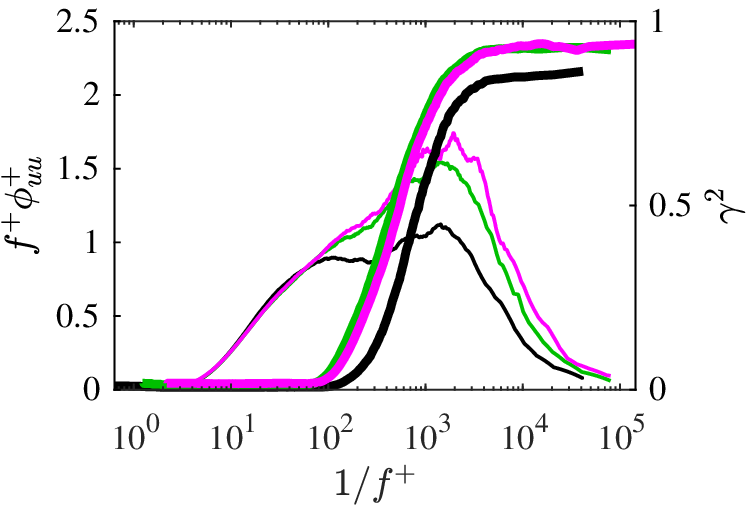}
\put(-360,126){(\textit{a})}
\put(-190,126){(\textit{b})}
\caption{(\textit{a}) $|\widetilde{H}_L|$, gain of the linear kernel and (\textit{b}) $\gamma^2$, linear coherence spectra between the fixed outer and moving inner probe at $z^+ =100$. Both quantities are filtered by a 25\% bandwidth moving filter. Line colours black, green and magenta represent the smooth, \RtoSLow{} and \RtoSHigh{} cases, respectively. The premultiplied energy spectrum $f^+\phi_{uu}^+$ is also shown in (\textit{b}) by thin lines of corresponding colours on the left vertical axis for reference.}
\label{fig:HL_gamma}
\end{figure}

The sensitivity of small scales to amplitude modulation is indicated by $\Gamma$: for a higher $\Gamma$, the universal small-scale signal $u^*$ will be multiplied by a higher fraction of the superposition signal $u_S^+$ to generate the prediction. Two calibrations of a smooth-wall turbulent boundary layer at $Re_{\tau}\approx 7350$ and 13300 result in very similar $\Gamma$ \citep{baars2016spectral}. For rough-to-smooth cases, as shown in Fig. \ref{fig:Gammavec}, $\Gamma$ from the three calibrations reach a good overall agreement, suggesting that the over-energised rough-wall structures in the outer layer do not seem to alter the amplitude modulation mechanism in the near-wall region.

Fig. \ref{fig:HL_gamma}(\textit{a}) shows $|\widetilde{H}_L|$, the gain of the linear kernel (which relates the superposition $u_S^+$ to the outer-layer large-scale signal $u_O^+$ via Eq. \ref{eq:HL}), at a wall-normal position of $z^+=100$. Similar trends are also seen in other wall-normal positions, and are not shown here for brevity. The magnitude of the linear transfer kernel $|\widetilde{H}_L|$ is found to increase in the rough-to-smooth cases compared to the smooth-wall reference. The linear coherence spectra are shown in Fig. \ref{fig:HL_gamma}(\textit{b}). In the rough-to-smooth cases, $\gamma^2$ deviates from 0 at a smaller $1/f^+$, and remains higher than that of the smooth-wall reference. A higher $\gamma^2$ indicates a stronger correlation between the velocity fluctuations obtained by the inner and outer probes. These results suggest that for a given structure in the outer layer, it will leave a stronger footprint (superposition) in the near-wall region in the rough-to-smooth case as a result of the enhanced inner-outer coherence. 

\subsection{Discussion} \label{subsec:case_discussion}

Overall, the amplitude modulation mechanism appears to be little modified after introducing a roughness heterogeneity (at least at $\hat{x}/\delta_0=2.3$ as examined here). However, there is a stronger correlation between the large-scale velocity fluctuations obtained by the inner and outer probes, and the gain in the linear kernel is also higher in rough-to-smooth cases. The large-scale fluctuation $\left<u_O^{+2}\right>$ is already stronger in the rough-to-smooth cases, and a larger fraction of it will contribute to the near-wall superposition signal through the increased gain $|\widetilde{H}_L|$. In light of the analysis in Sec. \ref{sec:R_IOIM}, the increase of $\left<u_S^{+2}\right>$ is primarily responsible for the higher $R$ value downstream of a rough-to-smooth change.

It is interesting to draw a direct comparison between the FST \citep{dogan2016interactions, dogan2017modelling} and the current rough-to-smooth cases, as both are in the `top-down' category with broadband energetic large-scale motions imposed in the outer region. The increased strength in the outer large scales makes them less susceptible to the interruptions from near-wall motions, leading to an increase in the correlation between outer- and near-wall large-scale signals, which is eventually reflected in an higher $|\widetilde{H}_L|$ in both. The major difference between the two is in the coefficient $\Gamma$: the former has a lower $\Gamma$, while the value is unchanged in the latter, similar to the independence of $\Gamma$ on $Re_{\tau}$ values observed in canonical smooth-wall boundary layers \citep{baars2016spectral}. We speculate that such difference is rooted in the manner in which the outer large scales are organised, as well as the energetic wavelengths. In the FST case, the outer structures are created by an active grid, which are inherently different from the structures organised by hairpin vortices in a developing boundary layer. The dissimilarity in the generation mechanism between large- and small-scale structures in the FST might be the reason for the reduction in $\Gamma$. The highly energetic large scales arranged in a manner similar to that of a naturally developed boundary layer in the current configuration makes it a good mimetic of a smooth-wall turbulent boundary layer at very high Reynolds numbers.

\section{Concluding remarks} \label{sec:conclusions}

In this work, we first show the underlying physics behind the increased amplitude modulation coefficient as reported in many previous studies, utilising the framework of IOIM. An analytical relationship between the amplitude modulation coefficient $R_2$ and IOIM parameters is derived and verified using a smooth-wall turbulent boundary layer dataset. This framework is then applied to classify and interpret the reported amplitude modulation behaviours in previous works. We then present the case study of a turbulent boundary layer downstream of a rough-to-smooth change with both single probe and simultaneous two-probe measurements. A stronger amplitude modulation effect evidenced by an increased $R$ is observed. Further analysis of the two-probe data reveals that the modulation strength $\Gamma$ is similar to that of a canonical smooth-wall turbulent boundary layer, and it remains the same for the two Reynolds numbers tested. The increase in $R$ is primarily attributed to the stronger large-scale footprints $\left<u_S^{+2}\right>$ in the near-wall region, which is contributed by both the over-energetic outer layer motions $\left<u_O^{+2}\right>$ and stronger coherence between the inner and outer layers. These results and analyses offer a new perspective to interpret the abundant literature on the the scale interactions of non-canonical turbulent boundary layers, which can be meaningful for incorporating the IOIM in the numerical simulation of a wide range of flow conditions.

\bibliography{MAIN_BIB}

\begin{thebibliography}{58}%
\makeatletter
\providecommand \@ifxundefined [1]{%
 \@ifx{#1\undefined}
}%
\providecommand \@ifnum [1]{%
 \ifnum #1\expandafter \@firstoftwo
 \else \expandafter \@secondoftwo
 \fi
}%
\providecommand \@ifx [1]{%
 \ifx #1\expandafter \@firstoftwo
 \else \expandafter \@secondoftwo
 \fi
}%
\providecommand \natexlab [1]{#1}%
\providecommand \enquote  [1]{``#1''}%
\providecommand \bibnamefont  [1]{#1}%
\providecommand \bibfnamefont [1]{#1}%
\providecommand \citenamefont [1]{#1}%
\providecommand \href@noop [0]{\@secondoftwo}%
\providecommand \href [0]{\begingroup \@sanitize@url \@href}%
\providecommand \@href[1]{\@@startlink{#1}\@@href}%
\providecommand \@@href[1]{\endgroup#1\@@endlink}%
\providecommand \@sanitize@url [0]{\catcode `\\12\catcode `\$12\catcode
  `\&12\catcode `\#12\catcode `\^12\catcode `\_12\catcode `\%12\relax}%
\providecommand \@@startlink[1]{}%
\providecommand \@@endlink[0]{}%
\providecommand \url  [0]{\begingroup\@sanitize@url \@url }%
\providecommand \@url [1]{\endgroup\@href {#1}{\urlprefix }}%
\providecommand \urlprefix  [0]{URL }%
\providecommand \Eprint [0]{\href }%
\providecommand \doibase [0]{https://doi.org/}%
\providecommand \selectlanguage [0]{\@gobble}%
\providecommand \bibinfo  [0]{\@secondoftwo}%
\providecommand \bibfield  [0]{\@secondoftwo}%
\providecommand \translation [1]{[#1]}%
\providecommand \BibitemOpen [0]{}%
\providecommand \bibitemStop [0]{}%
\providecommand \bibitemNoStop [0]{.\EOS\space}%
\providecommand \EOS [0]{\spacefactor3000\relax}%
\providecommand \BibitemShut  [1]{\csname bibitem#1\endcsname}%
\let\auto@bib@innerbib\@empty
\bibitem [{\citenamefont {Balakumar}\ and\ \citenamefont
  {Adrian}(2007)}]{balakumar2007large}%
  \BibitemOpen
  \bibfield  {author} {\bibinfo {author} {\bibfnamefont {B.~J.}\ \bibnamefont
  {Balakumar}}\ and\ \bibinfo {author} {\bibfnamefont {R.~J.}\ \bibnamefont
  {Adrian}},\ }\bibfield  {title} {\bibinfo {title} {Large- and
  very-large-scale motions in channel and boundary-layer flows},\ }\href@noop
  {} {\bibfield  {journal} {\bibinfo  {journal} {Phil. Trans. Roy. Soc. Lond.
  A: Math., Phys. Eng. Sci.}\ }\textbf {\bibinfo {volume} {365}},\ \bibinfo
  {pages} {665} (\bibinfo {year} {2007})}\BibitemShut {NoStop}%
\bibitem [{\citenamefont {Kim}\ and\ \citenamefont {Adrian}(1999)}]{Kim1999}%
  \BibitemOpen
  \bibfield  {author} {\bibinfo {author} {\bibfnamefont {K.~C.}\ \bibnamefont
  {Kim}}\ and\ \bibinfo {author} {\bibfnamefont {R.~J.}\ \bibnamefont
  {Adrian}},\ }\bibfield  {title} {\bibinfo {title} {Very large-scale motion in
  the outer layer},\ }\href@noop {} {\bibfield  {journal} {\bibinfo  {journal}
  {Phys. Fluids}\ }\textbf {\bibinfo {volume} {11}},\ \bibinfo {pages} {417}
  (\bibinfo {year} {1999})}\BibitemShut {NoStop}%
\bibitem [{\citenamefont {Smits}\ \emph {et~al.}(2011)\citenamefont {Smits},
  \citenamefont {McKeon},\ and\ \citenamefont {Marusic}}]{Smits2011a}%
  \BibitemOpen
  \bibfield  {author} {\bibinfo {author} {\bibfnamefont {A.~J.}\ \bibnamefont
  {Smits}}, \bibinfo {author} {\bibfnamefont {B.~J.}\ \bibnamefont {McKeon}},\
  and\ \bibinfo {author} {\bibfnamefont {I.}~\bibnamefont {Marusic}},\
  }\bibfield  {title} {\bibinfo {title} {High-{R}eynolds number wall
  turbulence},\ }\href@noop {} {\bibfield  {journal} {\bibinfo  {journal}
  {Annu. Rev. Fluid Mech.}\ }\textbf {\bibinfo {volume} {43}},\ \bibinfo
  {pages} {353} (\bibinfo {year} {2011})}\BibitemShut {NoStop}%
\bibitem [{\citenamefont {Hutchins}\ and\ \citenamefont
  {Marusic}(2007{\natexlab{a}})}]{Hutchins2007}%
  \BibitemOpen
  \bibfield  {author} {\bibinfo {author} {\bibfnamefont {N.}~\bibnamefont
  {Hutchins}}\ and\ \bibinfo {author} {\bibfnamefont {I.}~\bibnamefont
  {Marusic}},\ }\bibfield  {title} {\bibinfo {title} {Evidence of very long
  meandering features in the logarithmic region of turbulent boundary layers},\
  }\href@noop {} {\bibfield  {journal} {\bibinfo  {journal} {J. Fluid Mech}\
  }\textbf {\bibinfo {volume} {579}},\ \bibinfo {pages} {1} (\bibinfo {year}
  {2007}{\natexlab{a}})}\BibitemShut {NoStop}%
\bibitem [{\citenamefont {Hutchins}\ and\ \citenamefont
  {Marusic}(2007{\natexlab{b}})}]{Hutchins2007a}%
  \BibitemOpen
  \bibfield  {author} {\bibinfo {author} {\bibfnamefont {N.}~\bibnamefont
  {Hutchins}}\ and\ \bibinfo {author} {\bibfnamefont {I.}~\bibnamefont
  {Marusic}},\ }\bibfield  {title} {\bibinfo {title} {Large-scale influences in
  near-wall turbulence},\ }\href@noop {} {\bibfield  {journal} {\bibinfo
  {journal} {Phil. Trans. Royal Soc.}\ }\textbf {\bibinfo {volume} {365}},\
  \bibinfo {pages} {647} (\bibinfo {year} {2007}{\natexlab{b}})}\BibitemShut
  {NoStop}%
\bibitem [{\citenamefont {Hoyas}\ and\ \citenamefont
  {Jim{\'e}nez}(2006)}]{Hoyas2006}%
  \BibitemOpen
  \bibfield  {author} {\bibinfo {author} {\bibfnamefont {S.}~\bibnamefont
  {Hoyas}}\ and\ \bibinfo {author} {\bibfnamefont {J.}~\bibnamefont
  {Jim{\'e}nez}},\ }\bibfield  {title} {\bibinfo {title} {Scaling of the
  velocity fluctuations in turbulent channels up to ${R}e_\tau=2003$},\
  }\href@noop {} {\bibfield  {journal} {\bibinfo  {journal} {Phys. Fluids}\
  }\textbf {\bibinfo {volume} {18}},\ \bibinfo {pages} {011702} (\bibinfo
  {year} {2006})}\BibitemShut {NoStop}%
\bibitem [{\citenamefont {Vincenti}\ \emph {et~al.}(2013)\citenamefont
  {Vincenti}, \citenamefont {Klewicki}, \citenamefont {Morrill-Winter},
  \citenamefont {White},\ and\ \citenamefont
  {Wosnik}}]{vincenti2013streamwise}%
  \BibitemOpen
  \bibfield  {author} {\bibinfo {author} {\bibfnamefont {P.}~\bibnamefont
  {Vincenti}}, \bibinfo {author} {\bibfnamefont {J.}~\bibnamefont {Klewicki}},
  \bibinfo {author} {\bibfnamefont {C.}~\bibnamefont {Morrill-Winter}},
  \bibinfo {author} {\bibfnamefont {C.~M.}\ \bibnamefont {White}},\ and\
  \bibinfo {author} {\bibfnamefont {M.}~\bibnamefont {Wosnik}},\ }\bibfield
  {title} {\bibinfo {title} {Streamwise velocity statistics in turbulent
  boundary layers that spatially develop to high {R}eynolds number},\
  }\href@noop {} {\bibfield  {journal} {\bibinfo  {journal} {Exp. Fluids}\
  }\textbf {\bibinfo {volume} {54}},\ \bibinfo {pages} {1} (\bibinfo {year}
  {2013})}\BibitemShut {NoStop}%
\bibitem [{\citenamefont {Bandyopadhyay}\ and\ \citenamefont
  {Hussain}(1984)}]{bandyopadhyay1984coupling}%
  \BibitemOpen
  \bibfield  {author} {\bibinfo {author} {\bibfnamefont {P.~R.}\ \bibnamefont
  {Bandyopadhyay}}\ and\ \bibinfo {author} {\bibfnamefont {A.~K. M.~F.}\
  \bibnamefont {Hussain}},\ }\bibfield  {title} {\bibinfo {title} {The coupling
  between scales in shear flows},\ }\href@noop {} {\bibfield  {journal}
  {\bibinfo  {journal} {Phys. Fluids}\ }\textbf {\bibinfo {volume} {27}},\
  \bibinfo {pages} {2221} (\bibinfo {year} {1984})}\BibitemShut {NoStop}%
\bibitem [{\citenamefont {Agostini}\ and\ \citenamefont
  {Leschziner}(2014)}]{agostini2014influence}%
  \BibitemOpen
  \bibfield  {author} {\bibinfo {author} {\bibfnamefont {L.}~\bibnamefont
  {Agostini}}\ and\ \bibinfo {author} {\bibfnamefont {M.~A.}\ \bibnamefont
  {Leschziner}},\ }\bibfield  {title} {\bibinfo {title} {On the influence of
  outer large-scale structures on near-wall turbulence in channel flow},\
  }\href@noop {} {\bibfield  {journal} {\bibinfo  {journal} {Phys. Fluids}\
  }\textbf {\bibinfo {volume} {26}},\ \bibinfo {pages} {075107} (\bibinfo
  {year} {2014})}\BibitemShut {NoStop}%
\bibitem [{\citenamefont {He}(2018)}]{he2018multiscale}%
  \BibitemOpen
  \bibfield  {author} {\bibinfo {author} {\bibfnamefont {L.}~\bibnamefont
  {He}},\ }\bibfield  {title} {\bibinfo {title} {Multiscale block spectral
  solution for unsteady flows},\ }\href@noop {} {\bibfield  {journal} {\bibinfo
   {journal} {Int. J. Numer. Methods Fluids}\ }\textbf {\bibinfo {volume}
  {86}},\ \bibinfo {pages} {655} (\bibinfo {year} {2018})}\BibitemShut
  {NoStop}%
\bibitem [{\citenamefont {Chen}\ and\ \citenamefont {He}(2023)}]{chen2023two}%
  \BibitemOpen
  \bibfield  {author} {\bibinfo {author} {\bibfnamefont {C.}~\bibnamefont
  {Chen}}\ and\ \bibinfo {author} {\bibfnamefont {L.}~\bibnamefont {He}},\
  }\bibfield  {title} {\bibinfo {title} {Two-scale solution for tripped
  turbulent boundary layer},\ }\href@noop {} {\bibfield  {journal} {\bibinfo
  {journal} {J. Fluid Mech.}\ }\textbf {\bibinfo {volume} {955}},\ \bibinfo
  {pages} {A5} (\bibinfo {year} {2023})}\BibitemShut {NoStop}%
\bibitem [{\citenamefont {Zhang}\ and\ \citenamefont
  {Chernyshenko}(2016)}]{zhang2016quasisteady}%
  \BibitemOpen
  \bibfield  {author} {\bibinfo {author} {\bibfnamefont {C.}~\bibnamefont
  {Zhang}}\ and\ \bibinfo {author} {\bibfnamefont {S.~I.}\ \bibnamefont
  {Chernyshenko}},\ }\bibfield  {title} {\bibinfo {title} {Quasisteady
  quasihomogeneous description of the scale interactions in near-wall
  turbulence},\ }\href@noop {} {\bibfield  {journal} {\bibinfo  {journal}
  {Phys. Rev. Fluids}\ }\textbf {\bibinfo {volume} {1}},\ \bibinfo {pages}
  {014401} (\bibinfo {year} {2016})}\BibitemShut {NoStop}%
\bibitem [{\citenamefont {Chernyshenko}(2021)}]{chernyshenko2021extension}%
  \BibitemOpen
  \bibfield  {author} {\bibinfo {author} {\bibfnamefont {S.}~\bibnamefont
  {Chernyshenko}},\ }\bibfield  {title} {\bibinfo {title} {Extension of {QSQH}
  theory of scale interaction in near-wall turbulence to all velocity
  components},\ }\href@noop {} {\bibfield  {journal} {\bibinfo  {journal} {J.
  Fluid Mech.}\ }\textbf {\bibinfo {volume} {916}},\ \bibinfo {pages} {A52}
  (\bibinfo {year} {2021})}\BibitemShut {NoStop}%
\bibitem [{\citenamefont {Mathis}\ \emph
  {et~al.}(2009{\natexlab{a}})\citenamefont {Mathis}, \citenamefont
  {Hutchins},\ and\ \citenamefont {Marusic}}]{Mathis2009}%
  \BibitemOpen
  \bibfield  {author} {\bibinfo {author} {\bibfnamefont {R.}~\bibnamefont
  {Mathis}}, \bibinfo {author} {\bibfnamefont {N.}~\bibnamefont {Hutchins}},\
  and\ \bibinfo {author} {\bibfnamefont {I.}~\bibnamefont {Marusic}},\
  }\bibfield  {title} {\bibinfo {title} {Large-scale amplitude modulation of
  the small-scale structures in turbulent boundary layers},\ }\href@noop {}
  {\bibfield  {journal} {\bibinfo  {journal} {J. Fluid Mech.}\ }\textbf
  {\bibinfo {volume} {628}},\ \bibinfo {pages} {311} (\bibinfo {year}
  {2009}{\natexlab{a}})}\BibitemShut {NoStop}%
\bibitem [{\citenamefont {Chung}\ and\ \citenamefont
  {McKeon}(2010)}]{Chung2010}%
  \BibitemOpen
  \bibfield  {author} {\bibinfo {author} {\bibfnamefont {D.}~\bibnamefont
  {Chung}}\ and\ \bibinfo {author} {\bibfnamefont {B.~J.}\ \bibnamefont
  {McKeon}},\ }\bibfield  {title} {\bibinfo {title} {Large-eddy simulation of
  large-scale structures in long channel flow},\ }\href@noop {} {\bibfield
  {journal} {\bibinfo  {journal} {J. Fluid Mech.}\ }\textbf {\bibinfo {volume}
  {661}},\ \bibinfo {pages} {341} (\bibinfo {year} {2010})}\BibitemShut
  {NoStop}%
\bibitem [{\citenamefont {Guala}\ \emph {et~al.}(2011)\citenamefont {Guala},
  \citenamefont {Metzger},\ and\ \citenamefont
  {McKeon}}]{guala2011interactions}%
  \BibitemOpen
  \bibfield  {author} {\bibinfo {author} {\bibfnamefont {M.}~\bibnamefont
  {Guala}}, \bibinfo {author} {\bibfnamefont {M.}~\bibnamefont {Metzger}},\
  and\ \bibinfo {author} {\bibfnamefont {B.~J.}\ \bibnamefont {McKeon}},\
  }\bibfield  {title} {\bibinfo {title} {Interactions within the turbulent
  boundary layer at high {R}eynolds number},\ }\href@noop {} {\bibfield
  {journal} {\bibinfo  {journal} {J. Fluid Mech.}\ }\textbf {\bibinfo {volume}
  {666}},\ \bibinfo {pages} {573} (\bibinfo {year} {2011})}\BibitemShut
  {NoStop}%
\bibitem [{\citenamefont {Ganapathisubramani}\ \emph
  {et~al.}(2012)\citenamefont {Ganapathisubramani}, \citenamefont {Hutchins},
  \citenamefont {Monty}, \citenamefont {Chung},\ and\ \citenamefont
  {Marusic}}]{ganapathisubramani2012amplitude}%
  \BibitemOpen
  \bibfield  {author} {\bibinfo {author} {\bibfnamefont {B.}~\bibnamefont
  {Ganapathisubramani}}, \bibinfo {author} {\bibfnamefont {N.}~\bibnamefont
  {Hutchins}}, \bibinfo {author} {\bibfnamefont {J.~P.}\ \bibnamefont {Monty}},
  \bibinfo {author} {\bibfnamefont {D.}~\bibnamefont {Chung}},\ and\ \bibinfo
  {author} {\bibfnamefont {I.}~\bibnamefont {Marusic}},\ }\bibfield  {title}
  {\bibinfo {title} {Amplitude and frequency modulation in wall turbulence},\
  }\href@noop {} {\bibfield  {journal} {\bibinfo  {journal} {J. Fluid Mech.}\
  }\textbf {\bibinfo {volume} {712}},\ \bibinfo {pages} {61} (\bibinfo {year}
  {2012})}\BibitemShut {NoStop}%
\bibitem [{\citenamefont {Baars}\ \emph {et~al.}(2015)\citenamefont {Baars},
  \citenamefont {Talluru}, \citenamefont {Hutchins},\ and\ \citenamefont
  {Marusic}}]{baars2015wavelet}%
  \BibitemOpen
  \bibfield  {author} {\bibinfo {author} {\bibfnamefont {W.~J.}\ \bibnamefont
  {Baars}}, \bibinfo {author} {\bibfnamefont {K.~M.}\ \bibnamefont {Talluru}},
  \bibinfo {author} {\bibfnamefont {N.}~\bibnamefont {Hutchins}},\ and\
  \bibinfo {author} {\bibfnamefont {I.}~\bibnamefont {Marusic}},\ }\bibfield
  {title} {\bibinfo {title} {Wavelet analysis of wall turbulence to study
  large-scale modulation of small scales},\ }\href@noop {} {\bibfield
  {journal} {\bibinfo  {journal} {Exp. Fluids}\ }\textbf {\bibinfo {volume}
  {56}},\ \bibinfo {pages} {1} (\bibinfo {year} {2015})}\BibitemShut {NoStop}%
\bibitem [{\citenamefont {Marusic}\ \emph {et~al.}(2010)\citenamefont
  {Marusic}, \citenamefont {Mathis},\ and\ \citenamefont
  {Hutchins}}]{Marusic2010a}%
  \BibitemOpen
  \bibfield  {author} {\bibinfo {author} {\bibfnamefont {I.}~\bibnamefont
  {Marusic}}, \bibinfo {author} {\bibfnamefont {R.}~\bibnamefont {Mathis}},\
  and\ \bibinfo {author} {\bibfnamefont {N.}~\bibnamefont {Hutchins}},\
  }\bibfield  {title} {\bibinfo {title} {Predictive model for wall-bounded
  turbulent flow},\ }\href@noop {} {\bibfield  {journal} {\bibinfo  {journal}
  {Science}\ }\textbf {\bibinfo {volume} {329}},\ \bibinfo {pages} {193}
  (\bibinfo {year} {2010})}\BibitemShut {NoStop}%
\bibitem [{\citenamefont {Mathis}\ \emph {et~al.}(2011)\citenamefont {Mathis},
  \citenamefont {Hutchins},\ and\ \citenamefont {Marusic}}]{Mathis2011}%
  \BibitemOpen
  \bibfield  {author} {\bibinfo {author} {\bibfnamefont {R.}~\bibnamefont
  {Mathis}}, \bibinfo {author} {\bibfnamefont {N.}~\bibnamefont {Hutchins}},\
  and\ \bibinfo {author} {\bibfnamefont {I.}~\bibnamefont {Marusic}},\
  }\bibfield  {title} {\bibinfo {title} {A predictive inner--outer model for
  streamwise turbulence statistics in wall-bounded flows},\ }\href@noop {}
  {\bibfield  {journal} {\bibinfo  {journal} {J. Fluid Mech.}\ }\textbf
  {\bibinfo {volume} {681}},\ \bibinfo {pages} {537} (\bibinfo {year}
  {2011})}\BibitemShut {NoStop}%
\bibitem [{\citenamefont {Agostini}\ and\ \citenamefont
  {Leschziner}(2016)}]{agostini2016predicting}%
  \BibitemOpen
  \bibfield  {author} {\bibinfo {author} {\bibfnamefont {L.}~\bibnamefont
  {Agostini}}\ and\ \bibinfo {author} {\bibfnamefont {M.}~\bibnamefont
  {Leschziner}},\ }\bibfield  {title} {\bibinfo {title} {Predicting the
  response of small-scale near-wall turbulence to large-scale outer motions},\
  }\href@noop {} {\bibfield  {journal} {\bibinfo  {journal} {Phys. Fluids}\
  }\textbf {\bibinfo {volume} {28}},\ \bibinfo {pages} {015107} (\bibinfo
  {year} {2016})}\BibitemShut {NoStop}%
\bibitem [{\citenamefont {Baars}\ \emph {et~al.}(2016)\citenamefont {Baars},
  \citenamefont {Hutchins},\ and\ \citenamefont {Marusic}}]{baars2016spectral}%
  \BibitemOpen
  \bibfield  {author} {\bibinfo {author} {\bibfnamefont {W.~J.}\ \bibnamefont
  {Baars}}, \bibinfo {author} {\bibfnamefont {N.}~\bibnamefont {Hutchins}},\
  and\ \bibinfo {author} {\bibfnamefont {I.}~\bibnamefont {Marusic}},\
  }\bibfield  {title} {\bibinfo {title} {Spectral stochastic estimation of
  high-{R}eynolds-number wall-bounded turbulence for a refined inner-outer
  interaction model},\ }\href@noop {} {\bibfield  {journal} {\bibinfo
  {journal} {Phys. Rev. Fluids}\ }\textbf {\bibinfo {volume} {1}},\ \bibinfo
  {pages} {054406} (\bibinfo {year} {2016})}\BibitemShut {NoStop}%
\bibitem [{\citenamefont {Inoue}\ \emph {et~al.}(2012)\citenamefont {Inoue},
  \citenamefont {Mathis}, \citenamefont {Marusic},\ and\ \citenamefont
  {Pullin}}]{inoue2012inner}%
  \BibitemOpen
  \bibfield  {author} {\bibinfo {author} {\bibfnamefont {M.}~\bibnamefont
  {Inoue}}, \bibinfo {author} {\bibfnamefont {R.}~\bibnamefont {Mathis}},
  \bibinfo {author} {\bibfnamefont {I.}~\bibnamefont {Marusic}},\ and\ \bibinfo
  {author} {\bibfnamefont {D.~I.}\ \bibnamefont {Pullin}},\ }\bibfield  {title}
  {\bibinfo {title} {Inner-layer intensities for the flat-plate turbulent
  boundary layer combining a predictive wall-model with large-eddy
  simulations},\ }\href@noop {} {\bibfield  {journal} {\bibinfo  {journal}
  {Phys. Fluids}\ }\textbf {\bibinfo {volume} {24}},\ \bibinfo {pages} {075102}
  (\bibinfo {year} {2012})}\BibitemShut {NoStop}%
\bibitem [{\citenamefont {Cabrit}\ \emph {et~al.}(2014)\citenamefont {Cabrit},
  \citenamefont {Mathis}, \citenamefont {Kulandaivelu},\ and\ \citenamefont
  {Marusic}}]{cabrit2014towards}%
  \BibitemOpen
  \bibfield  {author} {\bibinfo {author} {\bibfnamefont {O.}~\bibnamefont
  {Cabrit}}, \bibinfo {author} {\bibfnamefont {R.}~\bibnamefont {Mathis}},
  \bibinfo {author} {\bibfnamefont {V.}~\bibnamefont {Kulandaivelu}},\ and\
  \bibinfo {author} {\bibfnamefont {I.}~\bibnamefont {Marusic}},\ }\bibfield
  {title} {\bibinfo {title} {Towards a statistically accurate wall-model for
  large-eddy simulation},\ }in\ \href@noop {} {\emph {\bibinfo {booktitle}
  {Proceedings of the $19^{th}$ Australasian Fluid Mechanics Conference}}}\
  (\bibinfo {year} {2014})\BibitemShut {NoStop}%
\bibitem [{\citenamefont {Sidebottom}\ \emph {et~al.}(2014)\citenamefont
  {Sidebottom}, \citenamefont {Cabrit}, \citenamefont {Marusic}, \citenamefont
  {Meneveau}, \citenamefont {Ooi},\ and\ \citenamefont
  {Jones}}]{sidebottom2014modelling}%
  \BibitemOpen
  \bibfield  {author} {\bibinfo {author} {\bibfnamefont {W.}~\bibnamefont
  {Sidebottom}}, \bibinfo {author} {\bibfnamefont {O.}~\bibnamefont {Cabrit}},
  \bibinfo {author} {\bibfnamefont {I.}~\bibnamefont {Marusic}}, \bibinfo
  {author} {\bibfnamefont {C.}~\bibnamefont {Meneveau}}, \bibinfo {author}
  {\bibfnamefont {A.}~\bibnamefont {Ooi}},\ and\ \bibinfo {author}
  {\bibfnamefont {D.}~\bibnamefont {Jones}},\ }\bibfield  {title} {\bibinfo
  {title} {Modelling of wall shear-stress fluctuations for large-eddy
  simulation},\ }in\ \href@noop {} {\emph {\bibinfo {booktitle} {Proceedings of
  the $19^{th}$ Australasian Fluid Mechanics Conference}}}\ (\bibinfo {year}
  {2014})\BibitemShut {NoStop}%
\bibitem [{\citenamefont {Mathis}\ \emph
  {et~al.}(2009{\natexlab{b}})\citenamefont {Mathis}, \citenamefont {Monty},
  \citenamefont {Hutchins},\ and\ \citenamefont
  {Marusic}}]{mathis2009comparison}%
  \BibitemOpen
  \bibfield  {author} {\bibinfo {author} {\bibfnamefont {R.}~\bibnamefont
  {Mathis}}, \bibinfo {author} {\bibfnamefont {J.~P.}\ \bibnamefont {Monty}},
  \bibinfo {author} {\bibfnamefont {N.}~\bibnamefont {Hutchins}},\ and\
  \bibinfo {author} {\bibfnamefont {I.}~\bibnamefont {Marusic}},\ }\bibfield
  {title} {\bibinfo {title} {Comparison of large-scale amplitude modulation in
  turbulent boundary layers, pipes, and channel flows},\ }\href@noop {}
  {\bibfield  {journal} {\bibinfo  {journal} {Phys. Fluids}\ }\textbf {\bibinfo
  {volume} {21}},\ \bibinfo {pages} {111703} (\bibinfo {year}
  {2009}{\natexlab{b}})}\BibitemShut {NoStop}%
\bibitem [{\citenamefont {Monty}\ \emph {et~al.}(2010)\citenamefont {Monty},
  \citenamefont {Chong}, \citenamefont {Mathis}, \citenamefont {Hutchins},
  \citenamefont {Marusic},\ and\ \citenamefont {Allen}}]{monty2010high}%
  \BibitemOpen
  \bibfield  {author} {\bibinfo {author} {\bibfnamefont {J.~P.}\ \bibnamefont
  {Monty}}, \bibinfo {author} {\bibfnamefont {M.~S.}\ \bibnamefont {Chong}},
  \bibinfo {author} {\bibfnamefont {R.}~\bibnamefont {Mathis}}, \bibinfo
  {author} {\bibfnamefont {N.}~\bibnamefont {Hutchins}}, \bibinfo {author}
  {\bibfnamefont {I.}~\bibnamefont {Marusic}},\ and\ \bibinfo {author}
  {\bibfnamefont {J.~J.}\ \bibnamefont {Allen}},\ }\bibfield  {title} {\bibinfo
  {title} {A high {R}eynolds number turbulent boundary layer with regular
  ‘{B}raille-type’ roughness},\ }in\ \href@noop {} {\emph {\bibinfo
  {booktitle} {IUTAM Symposium on The Physics of Wall-Bounded Turbulent Flows
  on Rough Walls}}}\ (\bibinfo {organization} {Springer},\ \bibinfo {year}
  {2010})\ pp.\ \bibinfo {pages} {69--75}\BibitemShut {NoStop}%
\bibitem [{\citenamefont {Squire}\ \emph
  {et~al.}(2016{\natexlab{a}})\citenamefont {Squire}, \citenamefont {Baars},
  \citenamefont {Hutchins},\ and\ \citenamefont {Marusic}}]{squire2016inner}%
  \BibitemOpen
  \bibfield  {author} {\bibinfo {author} {\bibfnamefont {D.~T.}\ \bibnamefont
  {Squire}}, \bibinfo {author} {\bibfnamefont {W.~J.}\ \bibnamefont {Baars}},
  \bibinfo {author} {\bibfnamefont {N.}~\bibnamefont {Hutchins}},\ and\
  \bibinfo {author} {\bibfnamefont {I.}~\bibnamefont {Marusic}},\ }\bibfield
  {title} {\bibinfo {title} {Inner--outer interactions in rough-wall
  turbulence},\ }\href@noop {} {\bibfield  {journal} {\bibinfo  {journal} {J.
  Turb.}\ }\textbf {\bibinfo {volume} {17}},\ \bibinfo {pages} {1159} (\bibinfo
  {year} {2016}{\natexlab{a}})}\BibitemShut {NoStop}%
\bibitem [{\citenamefont {Anderson}(2016)}]{anderson2016amplitude}%
  \BibitemOpen
  \bibfield  {author} {\bibinfo {author} {\bibfnamefont {W.}~\bibnamefont
  {Anderson}},\ }\bibfield  {title} {\bibinfo {title} {Amplitude modulation of
  streamwise velocity fluctuations in the roughness sublayer: evidence from
  large-eddy simulations},\ }\href@noop {} {\bibfield  {journal} {\bibinfo
  {journal} {J. Fluid Mech.}\ }\textbf {\bibinfo {volume} {789}},\ \bibinfo
  {pages} {567} (\bibinfo {year} {2016})}\BibitemShut {NoStop}%
\bibitem [{\citenamefont {Wu}\ \emph {et~al.}(2019)\citenamefont {Wu},
  \citenamefont {Christensen},\ and\ \citenamefont
  {Pantano}}]{wu2019modelling}%
  \BibitemOpen
  \bibfield  {author} {\bibinfo {author} {\bibfnamefont {S.}~\bibnamefont
  {Wu}}, \bibinfo {author} {\bibfnamefont {K.~T.}\ \bibnamefont
  {Christensen}},\ and\ \bibinfo {author} {\bibfnamefont {C.}~\bibnamefont
  {Pantano}},\ }\bibfield  {title} {\bibinfo {title} {Modelling smooth-and
  transitionally rough-wall turbulent channel flow by leveraging inner--outer
  interactions and principal component analysis},\ }\href@noop {} {\bibfield
  {journal} {\bibinfo  {journal} {J. Fluid Mech.}\ }\textbf {\bibinfo {volume}
  {863}},\ \bibinfo {pages} {407} (\bibinfo {year} {2019})}\BibitemShut
  {NoStop}%
\bibitem [{\citenamefont {Blackman}\ \emph {et~al.}(2019)\citenamefont
  {Blackman}, \citenamefont {Perret},\ and\ \citenamefont
  {Mathis}}]{blackman2019assessment}%
  \BibitemOpen
  \bibfield  {author} {\bibinfo {author} {\bibfnamefont {K.}~\bibnamefont
  {Blackman}}, \bibinfo {author} {\bibfnamefont {L.}~\bibnamefont {Perret}},\
  and\ \bibinfo {author} {\bibfnamefont {R.}~\bibnamefont {Mathis}},\
  }\bibfield  {title} {\bibinfo {title} {Assessment of inner--outer
  interactions in the urban boundary layer using a predictive model},\
  }\href@noop {} {\bibfield  {journal} {\bibinfo  {journal} {J. Fluid Mech.}\
  }\textbf {\bibinfo {volume} {875}},\ \bibinfo {pages} {44} (\bibinfo {year}
  {2019})}\BibitemShut {NoStop}%
\bibitem [{\citenamefont {Kim}\ \emph {et~al.}(2020)\citenamefont {Kim},
  \citenamefont {Blois}, \citenamefont {Best},\ and\ \citenamefont
  {Christensen}}]{kim2020experimental}%
  \BibitemOpen
  \bibfield  {author} {\bibinfo {author} {\bibfnamefont {T.}~\bibnamefont
  {Kim}}, \bibinfo {author} {\bibfnamefont {G.}~\bibnamefont {Blois}}, \bibinfo
  {author} {\bibfnamefont {J.~L.}\ \bibnamefont {Best}},\ and\ \bibinfo
  {author} {\bibfnamefont {K.~T.}\ \bibnamefont {Christensen}},\ }\bibfield
  {title} {\bibinfo {title} {Experimental evidence of amplitude modulation in
  permeable-wall turbulence},\ }\href@noop {} {\bibfield  {journal} {\bibinfo
  {journal} {J. Fluid Mech.}\ }\textbf {\bibinfo {volume} {887}} (\bibinfo
  {year} {2020})}\BibitemShut {NoStop}%
\bibitem [{\citenamefont {Khorasani}\ \emph {et~al.}(2022)\citenamefont
  {Khorasani}, \citenamefont {Luhar},\ and\ \citenamefont
  {Bagheri}}]{khorasani2022turbulent}%
  \BibitemOpen
  \bibfield  {author} {\bibinfo {author} {\bibfnamefont {S.~M.~H.}\
  \bibnamefont {Khorasani}}, \bibinfo {author} {\bibfnamefont {M.}~\bibnamefont
  {Luhar}},\ and\ \bibinfo {author} {\bibfnamefont {S.}~\bibnamefont
  {Bagheri}},\ }\bibfield  {title} {\bibinfo {title} {Turbulent flows over
  engineered anisotropic porous substrates},\ }\href@noop {} {\bibfield
  {journal} {\bibinfo  {journal} {arXiv preprint arXiv:2210.10140}\ } (\bibinfo
  {year} {2022})}\BibitemShut {NoStop}%
\bibitem [{\citenamefont {Dogan}\ \emph {et~al.}(2016)\citenamefont {Dogan},
  \citenamefont {Hanson},\ and\ \citenamefont
  {Ganapathisubramani}}]{dogan2016interactions}%
  \BibitemOpen
  \bibfield  {author} {\bibinfo {author} {\bibfnamefont {E.}~\bibnamefont
  {Dogan}}, \bibinfo {author} {\bibfnamefont {R.~E.}\ \bibnamefont {Hanson}},\
  and\ \bibinfo {author} {\bibfnamefont {B.}~\bibnamefont
  {Ganapathisubramani}},\ }\bibfield  {title} {\bibinfo {title} {Interactions
  of large-scale free-stream turbulence with turbulent boundary layers},\
  }\href@noop {} {\bibfield  {journal} {\bibinfo  {journal} {J. Fluid Mech.}\
  }\textbf {\bibinfo {volume} {802}},\ \bibinfo {pages} {79} (\bibinfo {year}
  {2016})}\BibitemShut {NoStop}%
\bibitem [{\citenamefont {Dogan}\ \emph {et~al.}(2017)\citenamefont {Dogan},
  \citenamefont {Hearst},\ and\ \citenamefont
  {Ganapathisubramani}}]{dogan2017modelling}%
  \BibitemOpen
  \bibfield  {author} {\bibinfo {author} {\bibfnamefont {E.}~\bibnamefont
  {Dogan}}, \bibinfo {author} {\bibfnamefont {R.~J.}\ \bibnamefont {Hearst}},\
  and\ \bibinfo {author} {\bibfnamefont {B.}~\bibnamefont
  {Ganapathisubramani}},\ }\bibfield  {title} {\bibinfo {title} {Modelling high
  {R}eynolds number wall--turbulence interactions in laboratory experiments
  using large-scale free-stream turbulence},\ }\href@noop {} {\bibfield
  {journal} {\bibinfo  {journal} {Phil. Trans. Roy. Soc. Lond. A: Math., Phys.
  Eng. Sci.}\ }\textbf {\bibinfo {volume} {375}},\ \bibinfo {pages} {20160091}
  (\bibinfo {year} {2017})}\BibitemShut {NoStop}%
\bibitem [{\citenamefont {Duvvuri}\ and\ \citenamefont
  {McKeon}(2015)}]{duvvuri2015triadic}%
  \BibitemOpen
  \bibfield  {author} {\bibinfo {author} {\bibfnamefont {S.}~\bibnamefont
  {Duvvuri}}\ and\ \bibinfo {author} {\bibfnamefont {B.~J.}\ \bibnamefont
  {McKeon}},\ }\bibfield  {title} {\bibinfo {title} {Triadic scale interactions
  in a turbulent boundary layer},\ }\href@noop {} {\bibfield  {journal}
  {\bibinfo  {journal} {J. Fluid Mech.}\ }\textbf {\bibinfo {volume} {767}},\
  \bibinfo {pages} {R4} (\bibinfo {year} {2015})}\BibitemShut {NoStop}%
\bibitem [{\citenamefont {Lozier}\ \emph {et~al.}(2022)\citenamefont {Lozier},
  \citenamefont {Thomas},\ and\ \citenamefont
  {Gordeyev}}]{lozier2022experimental}%
  \BibitemOpen
  \bibfield  {author} {\bibinfo {author} {\bibfnamefont {M.}~\bibnamefont
  {Lozier}}, \bibinfo {author} {\bibfnamefont {F.~O.}\ \bibnamefont {Thomas}},\
  and\ \bibinfo {author} {\bibfnamefont {S.}~\bibnamefont {Gordeyev}},\
  }\bibfield  {title} {\bibinfo {title} {Experimental studies of boundary layer
  dynamics via active manipulation of large-scale structures},\ }in\ \href@noop
  {} {\emph {\bibinfo {booktitle} {Proceedings of the $12^{th}$ Int. Symp.
  Turb. Shear Flow Phenom.}}}\ (\bibinfo {address} {Osaka, Japan},\ \bibinfo
  {year} {2022})\BibitemShut {NoStop}%
\bibitem [{\citenamefont {Li}\ \emph {et~al.}(2019)\citenamefont {Li},
  \citenamefont {de~Silva}, \citenamefont {Baidya}, \citenamefont {Rouhi},
  \citenamefont {Chung}, \citenamefont {Marusic},\ and\ \citenamefont
  {Hutchins}}]{MogengJFM2019}%
  \BibitemOpen
  \bibfield  {author} {\bibinfo {author} {\bibfnamefont {M.}~\bibnamefont
  {Li}}, \bibinfo {author} {\bibfnamefont {C.~M.}\ \bibnamefont {de~Silva}},
  \bibinfo {author} {\bibfnamefont {R.}~\bibnamefont {Baidya}}, \bibinfo
  {author} {\bibfnamefont {A.}~\bibnamefont {Rouhi}}, \bibinfo {author}
  {\bibfnamefont {D.}~\bibnamefont {Chung}}, \bibinfo {author} {\bibfnamefont
  {I.}~\bibnamefont {Marusic}},\ and\ \bibinfo {author} {\bibfnamefont
  {N.}~\bibnamefont {Hutchins}},\ }\bibfield  {title} {\bibinfo {title}
  {Recovery of the wall-shear stress to equilibrium flow conditions after a
  rough-to-smooth step-change in turbulent boundary layers},\ }\href@noop {}
  {\bibfield  {journal} {\bibinfo  {journal} {J. Fluid Mech.}\ }\textbf
  {\bibinfo {volume} {872}},\ \bibinfo {pages} {472} (\bibinfo {year}
  {2019})}\BibitemShut {NoStop}%
\bibitem [{\citenamefont {Garratt}(1990)}]{Garratt1990}%
  \BibitemOpen
  \bibfield  {author} {\bibinfo {author} {\bibfnamefont {J.~R.}\ \bibnamefont
  {Garratt}},\ }\bibfield  {title} {\bibinfo {title} {The internal boundary
  layer -- {A} review},\ }\href@noop {} {\bibfield  {journal} {\bibinfo
  {journal} {Boundary-Layer Meteorol.}\ }\textbf {\bibinfo {volume} {50}},\
  \bibinfo {pages} {171} (\bibinfo {year} {1990})}\BibitemShut {NoStop}%
\bibitem [{\citenamefont {Elliott}(1958)}]{Elliott1958}%
  \BibitemOpen
  \bibfield  {author} {\bibinfo {author} {\bibfnamefont {W.~P.}\ \bibnamefont
  {Elliott}},\ }\bibfield  {title} {\bibinfo {title} {The growth of the
  atmospheric internal boundary layer},\ }\href@noop {} {\bibfield  {journal}
  {\bibinfo  {journal} {Trans. Am. Geophys. Union}\ }\textbf {\bibinfo {volume}
  {39}},\ \bibinfo {pages} {1048} (\bibinfo {year} {1958})}\BibitemShut
  {NoStop}%
\bibitem [{\citenamefont {Antonia}\ and\ \citenamefont
  {Luxton}(1972)}]{antonia1972response}%
  \BibitemOpen
  \bibfield  {author} {\bibinfo {author} {\bibfnamefont {R.~A.}\ \bibnamefont
  {Antonia}}\ and\ \bibinfo {author} {\bibfnamefont {R.~E.}\ \bibnamefont
  {Luxton}},\ }\bibfield  {title} {\bibinfo {title} {The response of a
  turbulent boundary layer to a step change in surface roughness. {P}art 2.
  {R}ough-to-smooth},\ }\href@noop {} {\bibfield  {journal} {\bibinfo
  {journal} {J. Fluid Mech.}\ }\textbf {\bibinfo {volume} {53}},\ \bibinfo
  {pages} {737} (\bibinfo {year} {1972})}\BibitemShut {NoStop}%
\bibitem [{\citenamefont {Hanson}\ and\ \citenamefont
  {Ganapathisubramani}(2016)}]{Hanson2016}%
  \BibitemOpen
  \bibfield  {author} {\bibinfo {author} {\bibfnamefont {R.~E.}\ \bibnamefont
  {Hanson}}\ and\ \bibinfo {author} {\bibfnamefont {B.}~\bibnamefont
  {Ganapathisubramani}},\ }\bibfield  {title} {\bibinfo {title} {Development of
  turbulent boundary layers past a step change in wall roughness},\ }\href@noop
  {} {\bibfield  {journal} {\bibinfo  {journal} {J. Fluid Mech.}\ }\textbf
  {\bibinfo {volume} {795}},\ \bibinfo {pages} {494} (\bibinfo {year}
  {2016})}\BibitemShut {NoStop}%
\bibitem [{\citenamefont {Rouhi}\ \emph {et~al.}(2019)\citenamefont {Rouhi},
  \citenamefont {Chung},\ and\ \citenamefont {Hutchins}}]{rouhi2018}%
  \BibitemOpen
  \bibfield  {author} {\bibinfo {author} {\bibfnamefont {A.}~\bibnamefont
  {Rouhi}}, \bibinfo {author} {\bibfnamefont {D.}~\bibnamefont {Chung}},\ and\
  \bibinfo {author} {\bibfnamefont {N.}~\bibnamefont {Hutchins}},\ }\bibfield
  {title} {\bibinfo {title} {Direct numerical simulation of open channel flow
  over smooth-to-rough and rough-to-smooth step changes},\ }\href@noop {}
  {\bibfield  {journal} {\bibinfo  {journal} {J. Fluid Mech.}\ }\textbf
  {\bibinfo {volume} {866}},\ \bibinfo {pages} {450} (\bibinfo {year}
  {2019})}\BibitemShut {NoStop}%
\bibitem [{\citenamefont {Li}\ \emph {et~al.}(2021)\citenamefont {Li},
  \citenamefont {de~Silva}, \citenamefont {Chung}, \citenamefont {Pullin},
  \citenamefont {Marusic},\ and\ \citenamefont {Hutchins}}]{Mogengl2020}%
  \BibitemOpen
  \bibfield  {author} {\bibinfo {author} {\bibfnamefont {M.}~\bibnamefont
  {Li}}, \bibinfo {author} {\bibfnamefont {C.~M.}\ \bibnamefont {de~Silva}},
  \bibinfo {author} {\bibfnamefont {D.}~\bibnamefont {Chung}}, \bibinfo
  {author} {\bibfnamefont {D.~I.}\ \bibnamefont {Pullin}}, \bibinfo {author}
  {\bibfnamefont {I.}~\bibnamefont {Marusic}},\ and\ \bibinfo {author}
  {\bibfnamefont {N.}~\bibnamefont {Hutchins}},\ }\bibfield  {title} {\bibinfo
  {title} {Experimental study of a turbulent boundary layer with a
  rough-to-smooth change in surface conditions at high {R}eynolds numbers},\
  }\href@noop {} {\bibfield  {journal} {\bibinfo  {journal} {J. Fluid Mech.}\
  }\textbf {\bibinfo {volume} {923}},\ \bibinfo {pages} {A18} (\bibinfo {year}
  {2021})}\BibitemShut {NoStop}%
\bibitem [{\citenamefont {Li}\ \emph {et~al.}(2022)\citenamefont {Li},
  \citenamefont {de~Silva}, \citenamefont {Chung}, \citenamefont {Pullin},
  \citenamefont {Marusic},\ and\ \citenamefont
  {Hutchins}}]{mogeng2022modelling}%
  \BibitemOpen
  \bibfield  {author} {\bibinfo {author} {\bibfnamefont {M.}~\bibnamefont
  {Li}}, \bibinfo {author} {\bibfnamefont {C.~M.}\ \bibnamefont {de~Silva}},
  \bibinfo {author} {\bibfnamefont {D.}~\bibnamefont {Chung}}, \bibinfo
  {author} {\bibfnamefont {D.~I.}\ \bibnamefont {Pullin}}, \bibinfo {author}
  {\bibfnamefont {I.}~\bibnamefont {Marusic}},\ and\ \bibinfo {author}
  {\bibfnamefont {N.}~\bibnamefont {Hutchins}},\ }\bibfield  {title} {\bibinfo
  {title} {Modelling the downstream development of a turbulent boundary layer
  following a step change of roughness},\ }\href@noop {} {\bibfield  {journal}
  {\bibinfo  {journal} {J. Fluid Mech.}\ }\textbf {\bibinfo {volume} {949}},\
  \bibinfo {pages} {A7} (\bibinfo {year} {2022})}\BibitemShut {NoStop}%
\bibitem [{\citenamefont {Townsend}(1976)}]{Townsend1976}%
  \BibitemOpen
  \bibfield  {author} {\bibinfo {author} {\bibfnamefont {A.~A.}\ \bibnamefont
  {Townsend}},\ }\href@noop {} {\emph {\bibinfo {title} {The structure of
  turbulent shear flow}}},\ \bibinfo {edition} {2nd}\ ed.\ (\bibinfo
  {publisher} {Cambridge University Press.},\ \bibinfo {year}
  {1976})\BibitemShut {NoStop}%
\bibitem [{\citenamefont {Squire}\ \emph
  {et~al.}(2016{\natexlab{b}})\citenamefont {Squire}, \citenamefont
  {Morrill-Winter}, \citenamefont {Hutchins}, \citenamefont {Schultz},
  \citenamefont {Klewicki},\ and\ \citenamefont
  {Marusic}}]{squire2016comparison}%
  \BibitemOpen
  \bibfield  {author} {\bibinfo {author} {\bibfnamefont {D.~T.}\ \bibnamefont
  {Squire}}, \bibinfo {author} {\bibfnamefont {C.}~\bibnamefont
  {Morrill-Winter}}, \bibinfo {author} {\bibfnamefont {N.}~\bibnamefont
  {Hutchins}}, \bibinfo {author} {\bibfnamefont {M.~P.}\ \bibnamefont
  {Schultz}}, \bibinfo {author} {\bibfnamefont {J.~C.}\ \bibnamefont
  {Klewicki}},\ and\ \bibinfo {author} {\bibfnamefont {I.}~\bibnamefont
  {Marusic}},\ }\bibfield  {title} {\bibinfo {title} {Comparison of turbulent
  boundary layers over smooth and rough surfaces up to high {R}eynolds
  numbers},\ }\href@noop {} {\bibfield  {journal} {\bibinfo  {journal} {J.
  Fluid Mech.}\ }\textbf {\bibinfo {volume} {795}},\ \bibinfo {pages} {210}
  (\bibinfo {year} {2016}{\natexlab{b}})}\BibitemShut {NoStop}%
\bibitem [{\citenamefont {Marusic}\ \emph {et~al.}(2015)\citenamefont
  {Marusic}, \citenamefont {Chauhan}, \citenamefont {Kulandaivelu},\ and\
  \citenamefont {Hutchins}}]{marusic2015evolution}%
  \BibitemOpen
  \bibfield  {author} {\bibinfo {author} {\bibfnamefont {I.}~\bibnamefont
  {Marusic}}, \bibinfo {author} {\bibfnamefont {K.~A.}\ \bibnamefont
  {Chauhan}}, \bibinfo {author} {\bibfnamefont {V.}~\bibnamefont
  {Kulandaivelu}},\ and\ \bibinfo {author} {\bibfnamefont {N.}~\bibnamefont
  {Hutchins}},\ }\bibfield  {title} {\bibinfo {title} {Evolution of
  zero-pressure-gradient boundary layers from different tripping conditions},\
  }\href@noop {} {\bibfield  {journal} {\bibinfo  {journal} {J. Fluid Mech.}\
  }\textbf {\bibinfo {volume} {783}},\ \bibinfo {pages} {379} (\bibinfo {year}
  {2015})}\BibitemShut {NoStop}%
\bibitem [{\citenamefont {Bendat}\ and\ \citenamefont
  {Piersol}(2011)}]{bendat2011random}%
  \BibitemOpen
  \bibfield  {author} {\bibinfo {author} {\bibfnamefont {J.~S.}\ \bibnamefont
  {Bendat}}\ and\ \bibinfo {author} {\bibfnamefont {A.~G.}\ \bibnamefont
  {Piersol}},\ }\href@noop {} {\emph {\bibinfo {title} {Random data: analysis
  and measurement procedures}}}\ (\bibinfo  {publisher} {John Wiley \& Sons},\
  \bibinfo {year} {2011})\BibitemShut {NoStop}%
\bibitem [{\citenamefont {Baars}\ \emph {et~al.}(2017)\citenamefont {Baars},
  \citenamefont {Hutchins},\ and\ \citenamefont {Marusic}}]{baars2017reynolds}%
  \BibitemOpen
  \bibfield  {author} {\bibinfo {author} {\bibfnamefont {W.~J.}\ \bibnamefont
  {Baars}}, \bibinfo {author} {\bibfnamefont {N.}~\bibnamefont {Hutchins}},\
  and\ \bibinfo {author} {\bibfnamefont {I.}~\bibnamefont {Marusic}},\
  }\bibfield  {title} {\bibinfo {title} {{R}eynolds number trend of hierarchies
  and scale interactions in turbulent boundary layers},\ }\href@noop {}
  {\bibfield  {journal} {\bibinfo  {journal} {Phil. Trans. Roy. Soc. Lond. A:
  Math., Phys. Eng. Sci.}\ }\textbf {\bibinfo {volume} {375}},\ \bibinfo
  {pages} {20160077} (\bibinfo {year} {2017})}\BibitemShut {NoStop}%
\bibitem [{\citenamefont {Mathis}\ \emph {et~al.}(2013)\citenamefont {Mathis},
  \citenamefont {Marusic}, \citenamefont {Chernyshenko},\ and\ \citenamefont
  {Hutchins}}]{Mathis2013}%
  \BibitemOpen
  \bibfield  {author} {\bibinfo {author} {\bibfnamefont {R.}~\bibnamefont
  {Mathis}}, \bibinfo {author} {\bibfnamefont {I.}~\bibnamefont {Marusic}},
  \bibinfo {author} {\bibfnamefont {S.~I.}\ \bibnamefont {Chernyshenko}},\ and\
  \bibinfo {author} {\bibfnamefont {N.}~\bibnamefont {Hutchins}},\ }\bibfield
  {title} {\bibinfo {title} {Estimating wall-shear-stress fluctuations given an
  outer region input},\ }\href@noop {} {\bibfield  {journal} {\bibinfo
  {journal} {J. Fluid Mech.}\ }\textbf {\bibinfo {volume} {715}},\ \bibinfo
  {pages} {163} (\bibinfo {year} {2013})}\BibitemShut {NoStop}%
\bibitem [{\citenamefont {Pathikonda}\ and\ \citenamefont
  {Christensen}(2017)}]{pathikonda2017inner}%
  \BibitemOpen
  \bibfield  {author} {\bibinfo {author} {\bibfnamefont {G.}~\bibnamefont
  {Pathikonda}}\ and\ \bibinfo {author} {\bibfnamefont {K.~T.}\ \bibnamefont
  {Christensen}},\ }\bibfield  {title} {\bibinfo {title} {Inner--outer
  interactions in a turbulent boundary layer overlying complex roughness},\
  }\href@noop {} {\bibfield  {journal} {\bibinfo  {journal} {Phys. Rev.
  Fluids}\ }\textbf {\bibinfo {volume} {2}},\ \bibinfo {pages} {044603}
  (\bibinfo {year} {2017})}\BibitemShut {NoStop}%
\bibitem [{\citenamefont {Efstathiou}\ and\ \citenamefont
  {Luhar}(2018)}]{efstathiou2018mean}%
  \BibitemOpen
  \bibfield  {author} {\bibinfo {author} {\bibfnamefont {C.}~\bibnamefont
  {Efstathiou}}\ and\ \bibinfo {author} {\bibfnamefont {M.}~\bibnamefont
  {Luhar}},\ }\bibfield  {title} {\bibinfo {title} {Mean turbulence statistics
  in boundary layers over high-porosity foams},\ }\href@noop {} {\bibfield
  {journal} {\bibinfo  {journal} {J. Fluid Mech.}\ }\textbf {\bibinfo {volume}
  {841}},\ \bibinfo {pages} {351} (\bibinfo {year} {2018})}\BibitemShut
  {NoStop}%
\bibitem [{\citenamefont {Kulandaivelu}(2012)}]{Kulandaivelu2012}%
  \BibitemOpen
  \bibfield  {author} {\bibinfo {author} {\bibfnamefont {V.}~\bibnamefont
  {Kulandaivelu}},\ }\emph {\bibinfo {title} {Evolution of zero pressure
  gradient turbulent boundary layers from different initial conditions}},\
  \href@noop {} {Ph.D. thesis},\ \bibinfo  {school} {The University of
  Melbourne} (\bibinfo {year} {2012})\BibitemShut {NoStop}%
\bibitem [{\citenamefont {Ligrani}\ and\ \citenamefont
  {Bradshaw}(1987)}]{ligrani1987spatial}%
  \BibitemOpen
  \bibfield  {author} {\bibinfo {author} {\bibfnamefont {P.~M.}\ \bibnamefont
  {Ligrani}}\ and\ \bibinfo {author} {\bibfnamefont {P.}~\bibnamefont
  {Bradshaw}},\ }\bibfield  {title} {\bibinfo {title} {Spatial resolution and
  measurement of turbulence in the viscous sublayer using subminiature hot-wire
  probes},\ }\href@noop {} {\bibfield  {journal} {\bibinfo  {journal} {Exp.
  Fluids}\ }\textbf {\bibinfo {volume} {5}},\ \bibinfo {pages} {407} (\bibinfo
  {year} {1987})}\BibitemShut {NoStop}%
\bibitem [{\citenamefont {Jacobi}\ and\ \citenamefont
  {McKeon}(2013)}]{jacobi2013phase}%
  \BibitemOpen
  \bibfield  {author} {\bibinfo {author} {\bibfnamefont {I.}~\bibnamefont
  {Jacobi}}\ and\ \bibinfo {author} {\bibfnamefont {B.~J.}\ \bibnamefont
  {McKeon}},\ }\bibfield  {title} {\bibinfo {title} {Phase relationships
  between large and small scales in the turbulent boundary layer},\ }\href@noop
  {} {\bibfield  {journal} {\bibinfo  {journal} {Exp. Fluids}\ }\textbf
  {\bibinfo {volume} {54}},\ \bibinfo {pages} {1} (\bibinfo {year}
  {2013})}\BibitemShut {NoStop}%
\bibitem [{\citenamefont {Jacobi}\ and\ \citenamefont
  {McKeon}(2017)}]{jacobi2017phase}%
  \BibitemOpen
  \bibfield  {author} {\bibinfo {author} {\bibfnamefont {I.}~\bibnamefont
  {Jacobi}}\ and\ \bibinfo {author} {\bibfnamefont {B.~J.}\ \bibnamefont
  {McKeon}},\ }\bibfield  {title} {\bibinfo {title} {Phase-relationships
  between scales in the perturbed turbulent boundary layer},\ }\href@noop {}
  {\bibfield  {journal} {\bibinfo  {journal} {J. Turb.}\ }\textbf {\bibinfo
  {volume} {18}},\ \bibinfo {pages} {1120} (\bibinfo {year}
  {2017})}\BibitemShut {NoStop}%
\bibitem [{\citenamefont {Deshpande}\ \emph {et~al.}(2022)\citenamefont
  {Deshpande}, \citenamefont {Chandran}, \citenamefont {Smits},\ and\
  \citenamefont {Marusic}}]{deshpande2022relationship}%
  \BibitemOpen
  \bibfield  {author} {\bibinfo {author} {\bibfnamefont {R.}~\bibnamefont
  {Deshpande}}, \bibinfo {author} {\bibfnamefont {D.}~\bibnamefont {Chandran}},
  \bibinfo {author} {\bibfnamefont {A.~J.}\ \bibnamefont {Smits}},\ and\
  \bibinfo {author} {\bibfnamefont {I.}~\bibnamefont {Marusic}},\ }\bibfield
  {title} {\bibinfo {title} {The relationship between manipulated inter-scale
  phase and energy-efficient turbulent drag reduction},\ }\href@noop {}
  {\bibfield  {journal} {\bibinfo  {journal} {arXiv preprint arXiv:2211.07176}\
  } (\bibinfo {year} {2022})}\BibitemShut {NoStop}%
\end{thebibliography}%

\end{document}